\documentclass[iop]{emulateapj}

\usepackage{graphicx}
\usepackage{amsmath}  
\usepackage{amssymb} 
\usepackage{longtable}
\usepackage[lofdepth,lotdepth,caption=false]{subfig}
\usepackage{natbib}
\usepackage{multirow}
\usepackage{apjfonts}
\usepackage{epsf}
\usepackage{enumitem}

\newcommand{\nfehltthree}{10}
\newcommand{\nstars}{61}
\newcommand{\nmg}{10}
\newcommand{\nsi}{34}
\newcommand{\nca}{45}
\newcommand{\nti}{36}

\newcommand{\nufds}{8}

\newcommand{\fehmin}{$-$3.4}
\newcommand{\fehmax}{$-$1.1}
\newcommand{\alphafe}{\ensuremath{[\alpha/\rm{Fe}]}}
\newcommand{\alphafeavg}{\ensuremath{[\alpha/\rm{Fe}]_{\rm{avg}}}}
\newcommand{\alphafeatm}{\ensuremath{[\alpha/\rm{Fe}]_{\rm{atm}}}}
\newcommand{\xfe}{\ensuremath{[\alpha_{\rm{j}}/\rm{Fe}]}}

\newcommand{\like}{\ensuremath{\mathcal{L}}}

\newcommand{\Teff}{\ensuremath{T_{\rm{eff}}}} 
\newcommand{\logg}{\ensuremath{\log g}}

\newcommand{\tsnemin}{100}  
\def\spose#1{\hbox to 0pt{#1\hss}}
\def\simlt{\mathrel{\spose{\lower 3pt\hbox{$\mathchar"218$}}
     \raise 2.0pt\hbox{$\mathchar"13C$}}}
\def\simgt{\mathrel{\spose{\lower 3pt\hbox{$\mathchar"218$}}
     \raise 2.0pt\hbox{$\mathchar"13E$}}}

\shorttitle{Alpha Elements in UFDs}
\shortauthors{Vargas~et~al.}

\begin{document}

\title{The Distribution of Alpha Elements in Ultra-Faint Dwarf Galaxies}

\author{Luis C. Vargas\altaffilmark{1}} 
\author{Marla Geha\altaffilmark{1}}
\author{Evan N. Kirby\altaffilmark{2,4}}
\author{Joshua D. Simon\altaffilmark{3}}
\affil{\altaffilmark{1}Department of Astronomy, Yale University, 260 Whitney Ave., New Haven, CT~06511, USA; luis.vargas@yale.edu}
\affil{\altaffilmark{2}Department of Physics and Astronomy, University of California, Irvine, 4129 Reines Hall, Irvine, CA~92697, USA}
\affil{\altaffilmark{3}Observatories of the Carnegie Institution of Washington, 813 Santa Barbara St., Pasadena, CA~91101, USA}

\altaffiltext{4}{Center for Galaxy Evolution Fellow}

\begin{abstract}

The Milky Way ultra$-$faint dwarf galaxies (UFDs) contain some of the
oldest, most metal$-$poor stars in the Universe.
We present [Mg/Fe], [Si/Fe], [Ca/Fe], [Ti/Fe], and mean \alphafe\,
abundance ratios for \nstars\, individual red giant branch stars
across \nufds\, UFDs. This is the largest sample of alpha abundances
published to date in galaxies with absolute magnitudes \mbox{M$_{V}$
  $>$ $-$8}, including the first measurements for Segue~1, Canes
Venatici~II, Ursa Major~I, and Leo~T.  Abundances were determined via
medium$-$resolution Keck/DEIMOS spectroscopy and spectral
synthesis. The sample spans the metallicity range
\mbox{\fehmin\, $< \rm{[Fe/H]} <$ \fehmax}.  With the possible exception of
Segue~1 and Ursa~Major~II, the individual UFDs show \textit{on
average} lower \alphafe\, at higher metallicities, consistent with
enrichment from Type~Ia supernovae. Thus even the faintest galaxies
have undergone at least a limited level of chemical
self$-$enrichment.  Together with recent photometric studies, this
suggests that star formation in the UFDs was not a single burst, but
instead lasted at least as much as the minimum time delay of the onset
of Type~Ia supernovae ($\sim\tsnemin$ Myr) and less than $\sim{2}$
Gyr. We further show that the combined population of UFDs has an \alphafe\,
abundance pattern that is inconsistent with a flat, Galactic halo$-$like alpha
abundance trend, and is also qualitatively different from that of the
more luminous CVn~I dSph, which does show a hint of a plateau at very
low [Fe/H].
\end{abstract}


\section{Introduction}\label{intro_sec}

Ultra-faint dwarf (UFD) galaxies are the least luminous (M$_{V}>-8$)
known galaxies in the Universe (\citealt{Willman05a,Willman05b};
\citealt{Belokurov06a,Belokurov07a};
\citealt{Zucker06a,Zucker06b}; \citealt{Sakamoto06},
\citealt{Walsh07}; \citealt{Irwin07}). Spectroscopic observations 
of individual stars demonstrate that UFD galaxies are dark matter 
dominated \citep{Simon07}. They obey the metallicity$-$luminosity 
relation found in the brighter, classical
spheroidals (dSphs), and have large internal metallicity 
spreads greater than 0.5\,dex \citep{Kirby08b}. 

Recent HST photometry extending below the main sequence
turn$-$off demonstrates that at least three UFDs (Hercules, 
Ursa Major I and Leo IV) are composed exclusively of ancient 
stars \mbox{$\sim{13}$ Gyr} old \citep{Brown12}.  These data
further suggest that the star formation lasted for less than
\mbox{$\sim{2}$ Gyr.}  In spite of this small age spread, the large
metallicity spread in UFDs is indicative of a complex formation history. 
Metallicity spreads can arise in different ways. Star formation
in a UFD may proceed continuously or in bursts within a single halo, 
\textit{on average} increasing its metallicity over time 
\citep[e.g.,][]{Lanfranchi04,Revaz09}. Inhomogeneous gas mixing 
can also lead to a wide range of stellar metallicities within a single 
satellite \citep[e.g.,][]{Argast00,Oey00}. Finally, the merger of multiple 
progenitors with different mean metallicities may also produce a wide 
metallicity spread, as seen in recent simulations of more massive 
satellites \citep{Wise12a}. Determining more detailed abundances 
of stars in the UFDs will provide insight into the history of star formation 
at these very early epochs.

The \alphafe\,\footnote[1]{We reserve the use of unsubscripted \alphafe\, to
refer to alpha abundance ratios \textit{in general};  individual 
alpha elements are introduced where appropriate. For any elements 
A and B, we use the standard notation \mbox{[A/B]~$\equiv \rm{log}_{10}\left(\rm{N_{A}}/\rm{N_{A,\odot}}\right)$-$
\rm{log}_{10}\left(\rm{N_{B}}/\rm{N_{B,\odot}}\right)$.}} 
abundance ratios, including [Mg/Fe], [Si/Fe], [Ca/Fe], and [Ti/Fe], provide 
important constraints on the chemical evolution history of a stellar population.
In the most metal$-$poor stars, the ISM is polluted by the products of 
massive stellar evolution and core-collapse Type II supernovae (SNe).
The chemical yields from these explosions (\citealt{Woosley95}; \citealt{Nomoto06})
result in super$-$solar \alphafe\, values. These yields may depend on
the mass, metallicity, and explosion energy of the supernova.  Hence, 
individual Type II SNe may leave a unique signature in the observed 
abundance patterns, provided that (a) the gas did not have sufficient time 
to mix prior to the formation of the next generation of stars, or (b) the 
number of SNe was small, leading to stochastic sampling of the IMF. 
This would lead to intrinsic scatter in the [$\alpha$/Fe] ratios and/or 
abnormal abundance ratios.  Given their low average metallicities, the UFDs 
are one of the best places to search for the signature of chemical 
enrichment from metal$-$free Population III stars \citep{Frebel12}.

Type~Ia SNe output negligible amounts of alpha$-$elements 
in contrast to iron$-$peak elements resulting in lower \alphafe\, 
with rising [Fe/H]. Due to the time delay in the onset of Type~Ia SNe,
the low \alphafe\, signature is indicative of star formation lasting 
longer than the minimum time delay, \mbox{$t_{\rm{min,Ia}}\sim\tsnemin$ Myr} \citep{Totani08,Maoz12b}. 
The [Fe/H] at which \alphafe\, starts to decrease helps 
constrain the efficiency of star formation (\citealt{Pagel09} and 
references therein).  It thus provides a means to distinguish 
stellar populations with different origins. Spectroscopic studies 
of classical dSphs \citep[e.g.,][]{Shetrone01,Venn04,Kirby11b} 
reported significantly lower \alphafe\, in comparison to the 
observable Milky Way halo at \mbox{[Fe/H]\,$\gtrsim{-2.5}$}. 
This result has been used to show that the classical dSphs had 
a different chemical evolution than the progenitor(s) of the bulk 
of the inner Milky Way halo, which were likely more massive 
dwarf systems \citep{Robertson05}. In contrast to the inner halo
pattern, \citet{Nissen10} have reported a population of nearby, low \alphafe\, 
stars consistent with outer halo membership based on their kinematics, 
thus providing some indication that accreted systems with 
low \alphafe\, abundance ratios were important contributors 
to the outer halo. 

Our knowledge of the distribution of chemical abundances in UFDs, 
their chemical evolution, and their similarity/difference with the halo
stars, is still limited. High-resolution (R $\gtrsim$ 20,000) abundance 
studies have begun to address these issues in the UFDs by targeting 
the brightest RGB stars for abundance analysis. 
These include studies of Ursa Major\,II and 
Coma Berenices (\citealt{Frebel10b}, 
3 stars in each galaxy), Segue~1 (\citealt{Norris10a}, 1 star), Leo~IV 
(\citealt{Simon10}, 1 star), Bo\"{o}tes~I (\citealt{Feltzing09}; 7 stars; 
\citealt{Norris10b}, 1 star; \citealt{Gilmore13}, 7 stars), and Hercules 
(\citealt{Koch08}, 2 stars; \citealt{Aden11}, 11 stars). These studies have 
primarily targeted the \mbox{[Fe/H] $< -$2.0} regime. \citet{Aden11} 
reported decreasing [Ca/Fe] with [Fe/H] in a sample 
of 9 stars in Hercules with \mbox{[Fe/H] $< -2.0$}. 
In contrast, \citet{Frebel10b} found similar \alphafe\, 
abundance patterns at $\rm{[Fe/H]} < -2.5$ between the Coma~Berenices 
and Ursa~Major~II UFDs, and the (flat) \alphafe$-$[Fe/H] pattern in the inner 
Milky Way halo. Thus, the role of the UFDs in building even the
most metal$-$poor end of the inner halo is still unclear. 

High$-$resolution abundance studies of UFDs are currently 
limited to relatively bright stars with apparent magnitude 
$V\lesssim{19}$. Coupled with the sparseness of the RGBs 
in UFD systems, high$-$resolution abundance studies using  
8$-$10 meter class telescopes remain impractical for large 
samples. For example, the faintest star in a UFD studied to 
date at high$-$resolution is the \textit{brightest} known RGB star in 
Leo\,IV with an apparent magnitude of $V\sim{19.2}$ 
\citep{Simon10}.   In order to build statistically meaningful samples 
of abundance measurements, we turned to medium-resolution 
spectroscopy.  

Medium$-$resolution studies (R $\sim$ 2,000 $-$10,000) 
have recently begun to play a major role in obtaining precise abundances 
for larger stellar samples in both classical dSphs and UFDs.  While lower
spectral resolution reduces the number of chemical species available
for study, medium$-$resolution spectroscopy has been used to 
successfully measure both iron \citep{Allende06,Lee08,Kirby08a} 
and alpha$-$element abundances \citep{Kirby09,Lee11}. 
\citet{Kirby10} presented homogeneous Keck/DEIMOS medium$-$resolution 
abundances for thousands of stars in eight of the classical dSphs, showing 
that these systems may share a common trend of rising [$\alpha$/Fe] with 
decreasing [Fe/H] down to [Fe/H] $\sim{-2.5}$. \citet{Lai11} reported 
halo$-$like [$\alpha$/Fe] ratios in \mbox{B\"{o}otes I} spanning
\mbox{$-3.8\lesssim{}\rm{[Fe/H]}\lesssim{-1.5}$}, and \citet{Norris10c}
presented [C/Fe] for 16 stars in \mbox{B\"{o}otes I} and 3 stars in
\mbox{Segue 1}, showing a wide range of carbon enhancements.

In this paper, we present the first homogeneous abundances for
[Mg/Fe], [Si/Fe], [Ca/Fe], and [Ti/Fe] for \nstars\, stars in \nufds\,
of the UFDs: Segue 1 (Seg\,1), Coma Berenices (Com~Ber), Ursa Major\,II
(UMa\,II), Ursa Major\,I (UMa\,I), Canes Venatici II (CVn\,II), Leo\,IV,
and Hercules (Herc).  Our observations and abundance measurement
technique are summarized in \S\,\ref{obs_sec}$-$\ref{analysis_sec}.  We present our
abundance results in \S\,\ref{results_sec_individual}$-$\ref{results_sec_all} 
and discuss their implications in \S\,\ref{disc_sec}.

\section{Observations and Sample Selection}\label{obs_sec}

We determine spectroscopic abundances for the sample of 
UFD stars first presented by \citet{Simon07}, hereafter SG07, \citet{Geha09}, 
and \citet{Simon11}, hereafter S11.  The sample was observed with the 
Keck/DEIMOS spectrograph \citep{Faber03} using the 1200 
line\,mm$^{-1}$ grating, which 
provided wavelength coverage between 6300 and 9100~\AA\, 
with a resolution of $\sim{1.3}$~\AA\, FWHM.  Spectra
were reduced using a modified version of the {\tt spec2d} software
pipeline (version~1.1.4) developed by the DEEP2 team
\citep{Newman12,Cooper12} optimized for stellar spectra (SG07).  The
final one-dimensional spectra include the random uncertainties per
pixel.  Radial velocities are measured by cross$-$correlating the 
science spectra with stellar templates, and are used in this work to 
shift the science spectra to the rest frame. 

We analyze only stars identified as UFD members by SG07 and S11.
These authors selected members on the basis of: (i) position in 
color$-$magnitude space relative to an M92 isochrone shifted to
the UFD distance; (ii) radial velocity within $\sim{3}\sigma$ of the 
systemic UFD velocity; (iii) \ion{Na}{1} $\lambda\lambda$8183,8195 
equivalent width $\lesssim{1}$~\AA, and (iv) a loose cut based on 
a \ion{Ca}{2} infrared triplet (CaT) estimate of the stellar metallicity. 
The \ion{Na}{1} criterion prevents contamination by disk dwarfs that 
share similar radial velocities and magnitudes as the UFD 
RGB stars. We refer the reader to SG07 for a detailed explanation of 
the data reduction and membership selection for each UFD.

\section{Abundance Analysis}\label{analysis_sec}

The metallicities\footnote[2]{Throughout this paper, we use metallicity and 
[Fe/H] interchangeably.} of stars in our sample have been previously
presented in \citet{Kirby08b} and \citet{Simon11}. Here, we measure for the first time
[Mg/Fe], [Si/Fe], [Ca/Fe], [Ti/Fe], and an overall \alphafe\, abundance ratio
using the spectral matching technique described in \citet{Kirby10} with an 
expanded error analysis accounting for asymmetric uncertainties in 
the abundance ratios.

\subsection{Spectral Grid \& Element Masks}\label{gridmask_ssec}

\begin{figure*}[tpb!]
\centering
\subfloat{
\includegraphics[width=.47\textwidth]{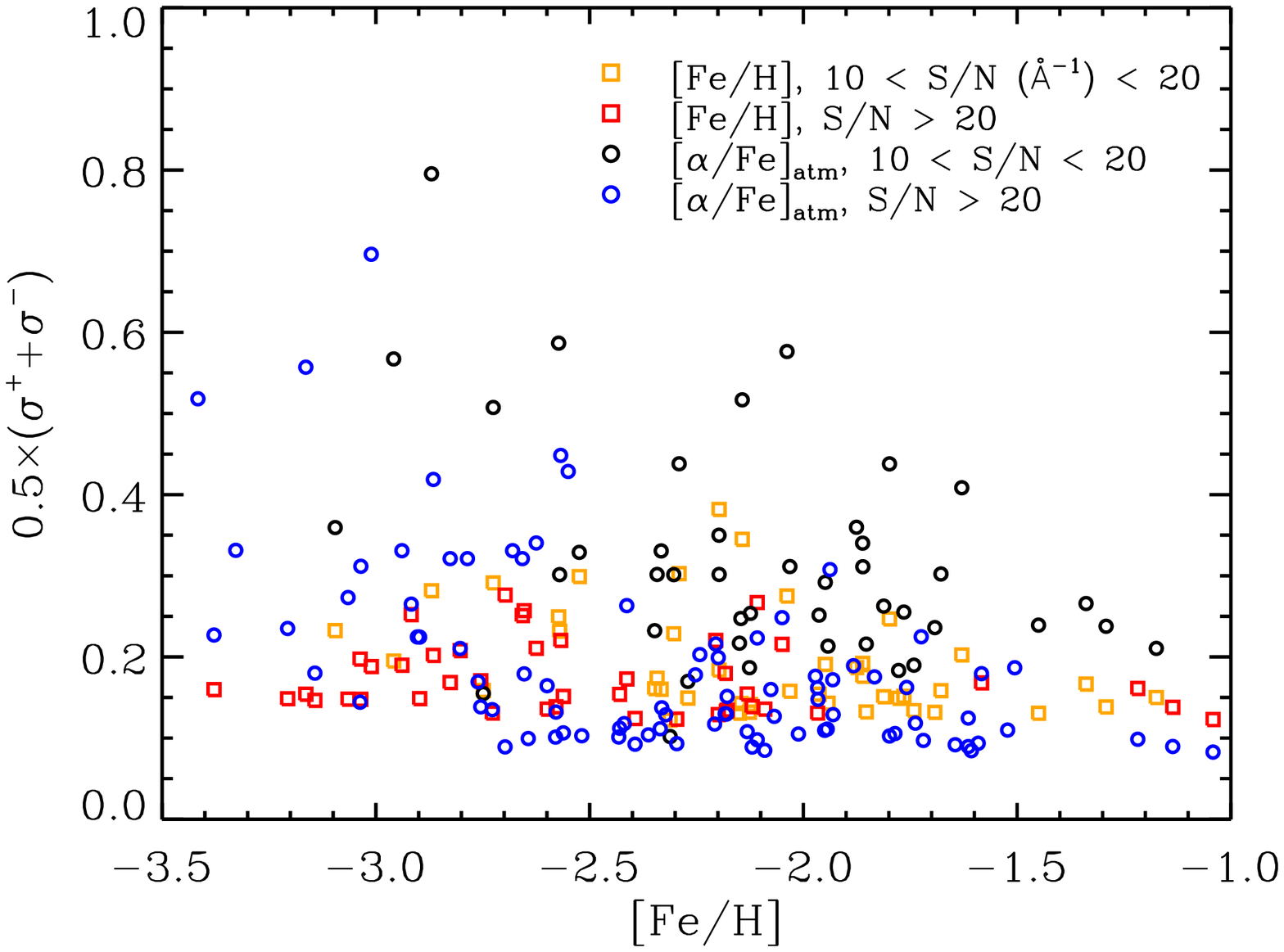}	
\label{subfig: erroraverage}}
\qquad	
\subfloat{
\includegraphics[width=.47\textwidth]{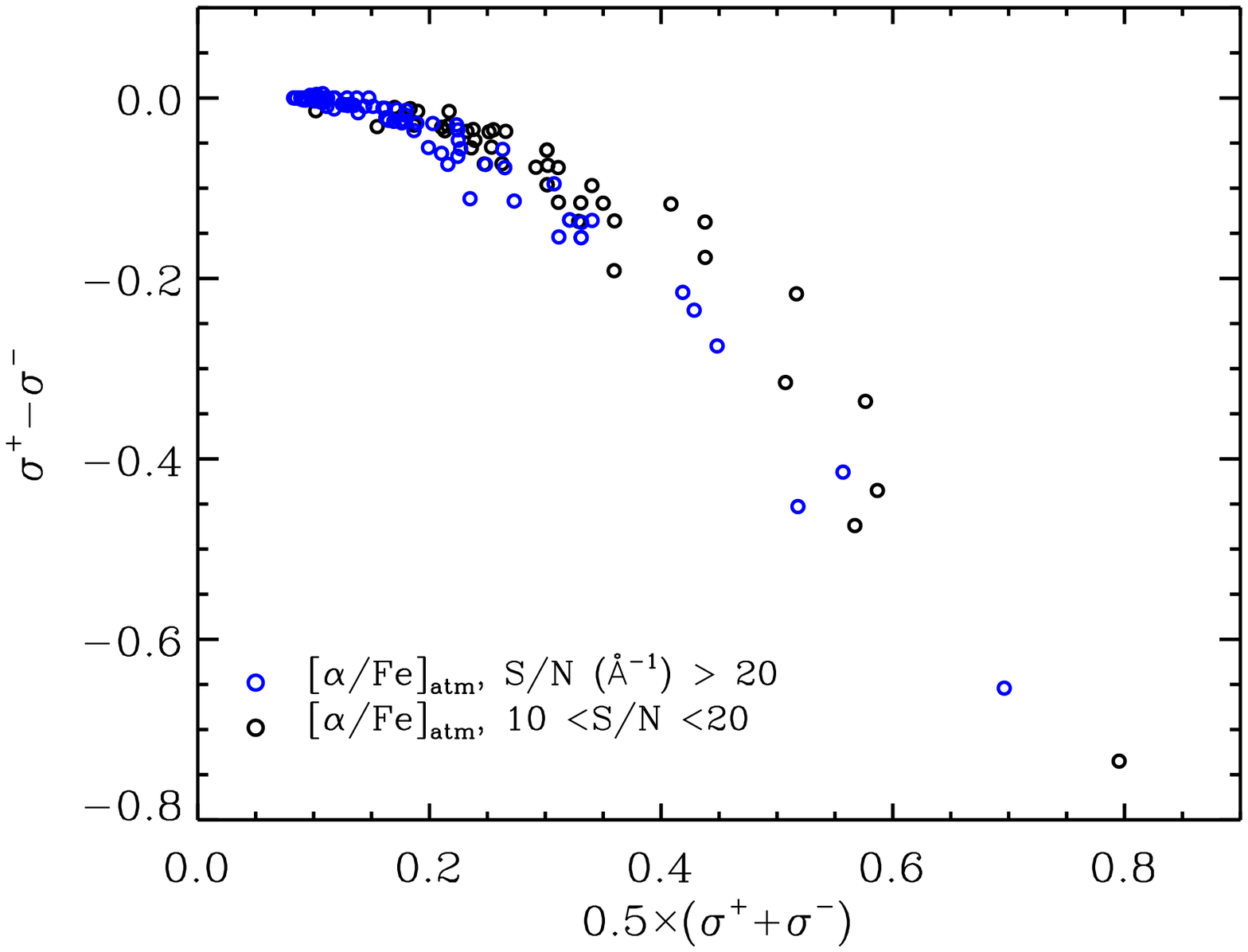}	
\label{subfig: errorasymmetry}}	
\caption{\textit{Left:} We plot the uncertainty in [Fe/H] and \alphafeatm\, 
as a function of [Fe/H] for stars with \mbox{S/N $>$ 20 \AA$^{-1}$} and 
\mbox{10 $<$ S/N $<$ 20 \AA$^{-1}$}. In the case of \alphafeatm, we
have averaged the $\sigma^{+}$ and $\sigma^{-}$ components for 
each measurement.  The colors and plot symbols denote uncertainties
in [Fe/H] or $\alphafeatm$ for different ranges of S/N. At a fixed S/N (\AA$^{-1}$), \alphafeatm\, 
uncertainties increase towards lower [Fe/H] due to progressively 
weaker features, whereas such a trend is not visible for [Fe/H], at
least for those stars with $\sigma_{\rm{[Fe/H]}} < 0.4$. The plot 
shows a minimum at $\sigma\sim{0.1}$ due to the element$-$dependent 
systematic uncertainty $\sigma_{\rm{sys}}$ added in quadrature to each 
random uncertainty. \textit{Right:} We show the asymmetry between 
uncertainties $\sigma^{+}_{\alphafeatm}$ and $\sigma^{-}_{\alphafeatm}$ for the same data, 
defined as $\sigma^{+}-\sigma^{-}$, plotted as a function of the mean 
uncertainty (average of $\sigma^{+}$ and $\sigma^{-}$). 
The negative component $\sigma^{-}$ is generally equal or larger 
than $\sigma^{+}$. A similar trend is present for the individual \xfe\, measurements.}
\label{fig: erroralphafefeh}
\end{figure*}

Our technique consists of a pixel$-$by$-$pixel matching 
between each stellar spectrum and a finely$-$spaced grid 
of synthetic spectra optimized for our spectral wavelength 
range. To measure stellar parameters, we rely on the 
synthetic spectral grid synthesized by \citet{Kirby11c} from 
plane$-$parallel ATLAS9 stellar atmospheres using the 
LTE abundance code MOOG \citep{Sneden73}. In addition,
we make use of an unpublished extension to the grid to
measure individual alpha abundance ratios, as described 
in \S\,\ref{abunds_ssec}. 

The primary synthetic spectral grid has four dimensions: \Teff, \logg, [Fe/H], 
and \alphafeatm. The quantity \alphafeatm\, 
is defined as the \alphafe\, abundance ratio of O, Ne, Mg, Si, S, 
Ar, Ca, and Ti used to synthesize each spectrum. The grid spans 
\mbox{3500 $\leq$ \Teff\,} $\leq$ 8000 K, \mbox{0.0 $\leq$ \logg\,}
$\leq$ 5.0, \mbox{$-$5.0 $\leq$ [Fe/H]} $\leq$ 0.0, 
and \mbox{$-$0.8 $\leq$ [$\alpha$/Fe]$_{\rm{atm}}$} $\leq$ +1.2. 
Our sample is comprised of RGB stars (\logg\, $<$ 3.6), making this
grid sufficient for our analysis. The microturbulent velocity, $\xi$, used 
for each synthesis was determined using an empirical $\xi-\logg$ relation 
valid for RGB stars, derived from high$-$resolution spectroscopic 
measurements (\citealt{Kirby09}, Equation 2). 

We perform our analysis using only spectral regions with Fe, Mg, Si, 
Ca, or Ti features to maximize sensitivity to each element. The 
mask of usable spectral regions for a given element X is constructed 
by synthesizing three spectra with \mbox{$\rm{[X/H]} = [-1.8,-1.5,-1.2]$}, 
while all other abundances remained fixed at $\rm{[X/H]} = -1.5$.
The mask is comprised of those wavelength segments where a 0.3 dex
difference in [X/H] changes the normalized flux by $\gtrsim{0.5}\%$. To 
incorporate regions sensitive at a wide range of \Teff\,, the procedure 
was repeated at 1,000~K intervals between 4,000~K and 8,000~K, 
and the resulting masks joined. The Mg, Si, Ca and Ti element masks 
do not share wavelength segments in common, allowing us to measure 
individual abundances in \S\,\ref{abunds_ssec}. The combined alpha mask is defined 
as the union of the Mg, Si, Ca, and Ti masks.
We remove from the element masks spectral lines that are not modeled 
accurately by the LTE synthesis code, as determined by \citet{Kirby08a} 
and listed in their Table 2. These include the \ion{Ca}{2} triplet and the 
\ion{Mg}{1} $\lambda{8807}$ feature. The Fe, Mg, Si, Ca, and Ti masks 
have roughly 222, 10, 20, 14, and 52 good segments each (i.e., not overlapping
with telluric regions or with badly modeled spectral lines), where each segment 
corresponds to a spectral feature. The combined spectral widths of 
the wavelength segments for each element are $\sim{500}$, 16, 20, 14, 
and 52 \AA, respectively\footnote{The number of segments will vary slightly 
from star to star due to slightly varying wavelength coverage and 
the presence of bad pixels and other imperfections in each spectrum.}.

\subsection{$\chi^{2}$ Pixel$-$by$-$Pixel Matching}\label{matching_ssec}

To perform the pixel fitting, we degrade the synthetic models to the 
DEIMOS spectral resolution. We account for a small quadratic 
dependence of the spectral FWHM on wavelength by fitting to unblended 
sky lines.  We then convolve the synthetic spectra with this variable 
FWHM Gaussian kernel. For each star, we determine \logg\, by 
fitting the SDSS photometry to a grid of Yale$-$Yonsei isochrones, 
as detailed in \citet{Kirby10}. The alternate spectroscopic approach, based
on obtaining ionization equilibrium between \ion{Fe}{1} and \ion{Fe}{2} 
abundances\footnote[3]{More generally, any element with two measurable
species, e.g. \ion{Ti}{1}$-$\ion{Ti}{2}; however, Fe by far contains the most signal.}, 
is not applicable to our data due to the dearth of absorption lines from ionized
species in our red spectra. We normalize the flux$-$calibrated 
spectrum using a low order spline fit to wavelength regions not 
sensitive to any of Fe, Mg, Si, Ca or Ti. The normalization is later 
refined during the fitting process. 

The best$-$fit parameters (\Teff, [Fe/H], and \alphafeatm)
and individual abundance ratios (\xfe, where \mbox{$\alpha_{j}$ = Mg}, Si, Ca and Ti in this work)
are determined by minimizing the $\chi^{2}$ statistic between the 
rest$-$frame science spectrum and the convolved model grid 
in a multi$-$step process described by \citet{Kirby10}. 
We briefly describe the fitting procedure for the various stellar parameters
and abundance ratios, highlighting the modifications implemented
for this paper. In particular, we have updated our uncertainty analysis 
to provide more accurate asymmetric \alphafeatm\, and \xfe\, uncertainties. 
\textit{Throughout, we maintain the order of steps described in detail by \citet{Kirby10}.}

\subsection{\Teff\, and [Fe/H]}\label{tefffeh_ssec}

We fit \Teff\, and [Fe/H] simultaneously using the 
Fe mask. Due to the wavelength overlap between the Fe 
and combined alpha masks, we do not fit [Fe/H] and 
\alphafeatm\, simultaneously. In order to optimize the fitting 
process in the two dimensional \Teff$-$[Fe/H] parameter 
space, we perform the $\chi^{2}$ minimization using 
the code \textit{mpfit} \citep{Markwardt09}, which is 
an IDL implementation of the Levenberg$-$Marquardt 
$\chi^{2}$ algorithm. 

We determine the random 
uncertainty in [Fe/H], \mbox{$\sigma_{\rm{[Fe/H],ran}}$}, by 
using the covariant error matrix of \Teff\, and [Fe/H] calculated
by \textit{mpfit}. Due to the non$-$zero cross$-$terms, 
\mbox{$\sigma_{\rm{[Fe/H],ran}}$} is larger than if the [Fe/H]
uncertainty was calculated by varying [Fe/H] alone. 
The total uncertainty in [Fe/H], \mbox{$\sigma_{\rm{[Fe/H]}}$}, is equal 
to the addition in quadrature of \mbox{$\sigma_{\rm{[Fe/H],ran}}$} 
to a systematic uncertainty component \mbox{$\sigma_{\rm{[Fe/H],sys}}$}.
\citet{Kirby10} estimated $\sigma_{\rm{[Fe/H],sys}}$ by 
calculating the residual difference between DEIMOS and 
high$-$resolution abundances of globular cluster 
stars, after accounting for the random uncertainty 
in both sets of measurements added in quadrature. 
In order to check the reliability of the \textit{mpfit}$-$derived uncertainties,
we calculate $\chi^{2}$ around the best$-$fit \Teff\, and [Fe/H]. We find
that $\chi^{2}$ contours for \Teff\, and [Fe/H] are symmetric
about the minimum $\chi^{2}$ value for \mbox{$\sigma_{\rm{[Fe/H]}}\lesssim{0.4}$},
justifying our use of the symmetric \textit{mpfit} random uncertainties. 
Henceforth, we only include stars with \mbox{$\sigma_{\rm{[Fe/H]}}\leq{0.4}$}.

\subsection{\alphafeatm\, and \xfe\, Abundance Ratios}\label{abunds_ssec}

We calculate \alphafeatm\, while fixing \Teff\, and 
[Fe/H] to the best$-$fit values, using the combined 
alpha mask defined in \S\,\ref{gridmask_ssec}.
We compute $\chi^{2}$ contours for \alphafeatm\,
by measuring the sum of the pixel$-$to$-$pixel 
variation between the stellar spectrum and the 
primary spectral grid. We measure the best$-$fit \alphafeatm\,
value by finding the value corresponding to the minimum
in the $\chi^{2}$ contour. The measurement of best$-$fit
\alphafeatm\, is analogous to that of Kirby et al., who performed
this optimization using \textit{mpfit}. After all
stellar parameters (\Teff, [Fe/H], and \alphafeatm) have
converged to their best$-$fit values, we fit for the
individual alpha abundances while keeping all
stellar parameters fixed.

To measure individual \xfe\, abundance ratios, 
we compare each spectrum to a supplementary spectral grid 
that samples values of \xfe\, from $-0.8$ to $+1.2$ dex, while 
keeping all other abundances and stellar parameters fixed. 
The grid was synthesized only for spectral regions included within 
each \xfe\, mask. We compute $\chi^{2}$ contours for each \xfe\, 
by measuring the \mbox{pixel$-$to$-$pixel} variation between 
each spectrum and the supplementary spectral grid, instead of 
the primary grid. 

In contrast to [Fe/H], we find that a significant number 
of \alphafeatm\, and \xfe\, contours are asymmetric 
about $\chi^{2}_{\rm{min}}$. We therefore estimate 
the random uncertainties by finding the two abundance 
values corresponding to \mbox{$\chi^{2}_{\rm{min}}$ + 1} 
without assuming symmetry. We refer to the positive and 
negative difference between these values and the best$-$fit 
abundance ratio, \alphafe\, as $\sigma^{+}_{\alphafe}$ and 
$\sigma^{-}_{\alphafe}$, respectively,
where \alphafe\, stands for any of \alphafeatm\, or \xfe.

We also account for non$-$random errors due to,
e.g., uncertainties in the other stellar parameters, by 
introducing a systematic error floor different for each
abundance ratio, $\sigma_{\alphafe\,\rm{sys}}$,
measured by \citet{Kirby10} in the same way as 
$\sigma_{\rm{[Fe/H],sys}}$ (\S\,\ref{tefffeh_ssec}).
The systematic uncertainties for \alphafeatm\, [Mg/Fe], [Si/Fe], [Ca/Fe] 
and [Ti/Fe] are 0.08, 0.11, 0.18, 0.09, and 0.10 dex, respectively.
We calculate the total uncertainty by adding 
$\sigma_{\alphafe\,\rm{sys}}$ to the 
$\sigma^{+}_{\alphafe\,\rm{ran}}$ and $\sigma^{+}_{\alphafe\,\rm{ran}}$ 
random components in quadrature. We note that 
$\sigma_{\alphafe\,\rm{sys}}$ only contributes significantly 
to the error budget when the random uncertainty is 
$\lesssim{0.1}$ dex. 

All abundances are referenced to the \citet{Asplund09} 
solar abundance scale. The offsets between the abundance 
scale used by \citet{Kirby08b,Kirby10} and this work are minimal: 
$+0.02$, $-0.04$, $+0.02$, $+0.00$, $+0.02$ dex for [Fe/H], 
[Mg/Fe], [Si/Fe], [Ca/Fe], and [Ti/Fe], respectively, in the 
sense of this work minus Kirby et al. There is no difference 
in the mean \alphafeatm\, between the old and new abundance 
scales.

\begin{figure}[t]
\centering
\includegraphics[width=0.47\textwidth]{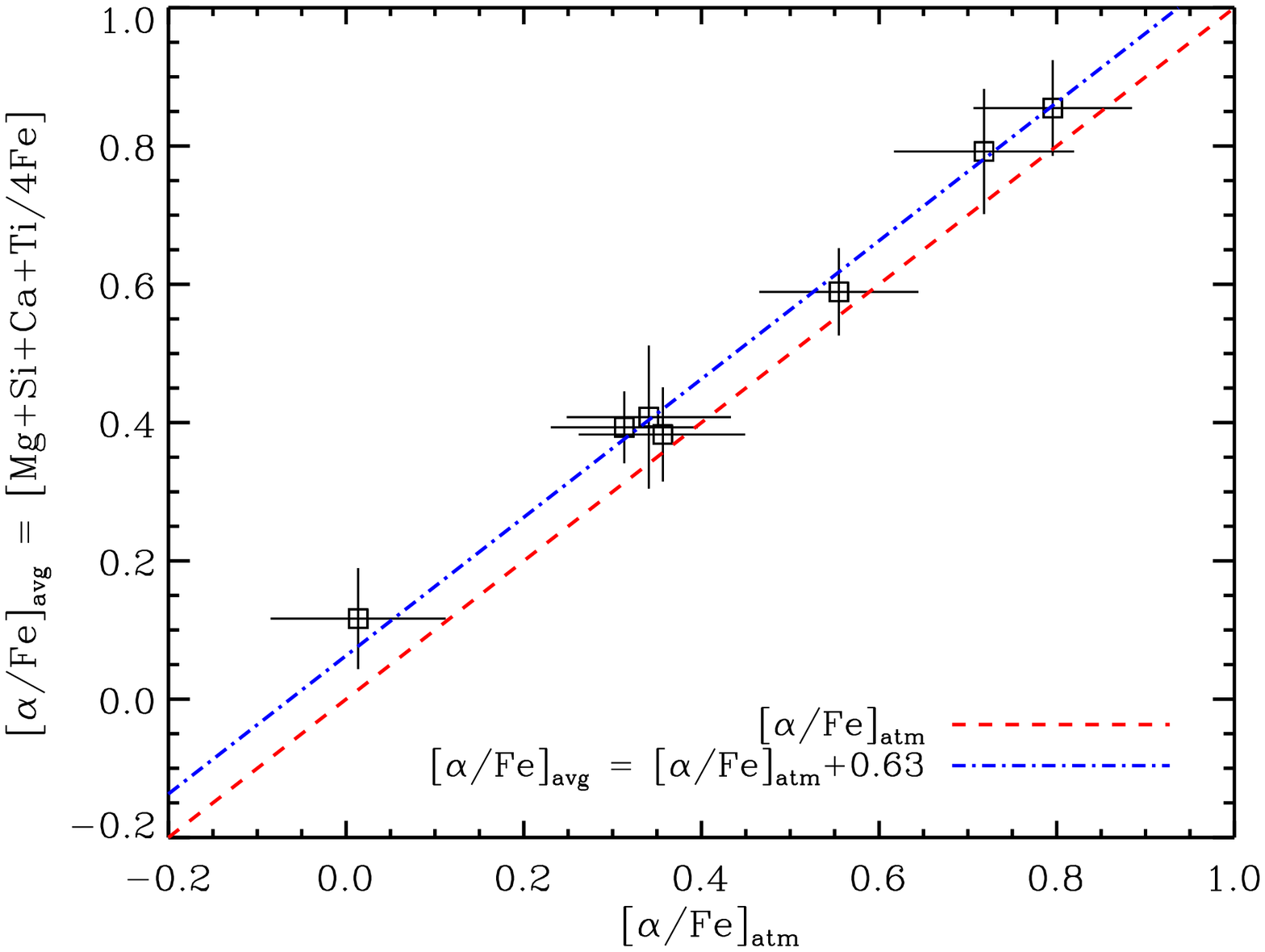}
\caption{We plot [$\alpha$/Fe]$_{\rm{atm}}$ against \alphafeavg\,
the mean of [Mg/Fe], [Si/Fe], [Ca/Fe], [Ti/Fe] for the nine stars 
with measurements available for all elements. The dashed 
(red) line is the $y=x$ line. The small offset between the data 
and the red line suggests a small difference between \alphafeatm\, 
\alphafeavg. The mean vertical offset is +0.063 dex. The dot$-$dash (blue) line shows 
the improved agreement obtained after adding the offset to \alphafeatm. 
We thus define \alphafeavg\, as \mbox{\alphafeatm\,$+ 0.63$}, and apply 
this correction to all stars in our sample.}
\label{fig: alphaatm}
\end{figure}

We show the uncertainty in [Fe/H] and \alphafeatm\, 
for all UFD stars in the left panel of Figure~\ref{fig: erroralphafefeh}. 
At a fixed S/N, the uncertainty increases towards lower 
[Fe/H] due to progressively weaker spectral features. 
The right panel shows the associated asymmetry in 
the \alphafeatm\, uncertainty as a function of its average
value. We find that $\sigma^{-}_{\alphafeatm}$ 
is preferentially larger than $\sigma^{+}_{\alphafeatm}$,
with a similar effect present for each \xfe\, (not shown in the figure). 
In our analysis, we include only abundances with 
\mbox{$\sigma_{\rm{[Fe/H]}}\leq{0.4}$} and 
\mbox{$\sigma^{+}_{\alphafeatm}$ $\leq$ 0.4 dex}
(\mbox{$\sigma^{+}_{\xfe}$ $\leq$ 0.4 dex}.)
The final sample includes \nstars, \nmg, \nsi, \nca, and \nti\, 
measurements of \alphafeatm, [Mg/Fe], [Si/Fe], [Ca/Fe], and [Ti/Fe], 
respectively.

In addition to the individual \xfe\,, we report an \textit{overall} 
alpha abundance ratio, which we denote as \alphafeavg. There
is no homogeneous definition of \alphafeavg\, in the literature. 
Different authors use different combinations of \xfe\, 
to estimate \alphafeavg. We choose [$\alpha$/Fe]$_{\rm{atm}}$ 
as our initial estimate of \alphafeavg\, because it was measured 
using the combined alpha mask, thus being sensitive 
to Mg, Si, Ca and Ti. \alphafeatm\,
has the added advantage of being measurable even 
when individual \xfe\, are not, because it is measured 
from the combined signal of four elements. 

\begin{figure*}[t!]
\centering
\subfloat{
\includegraphics[width=.47\textwidth]{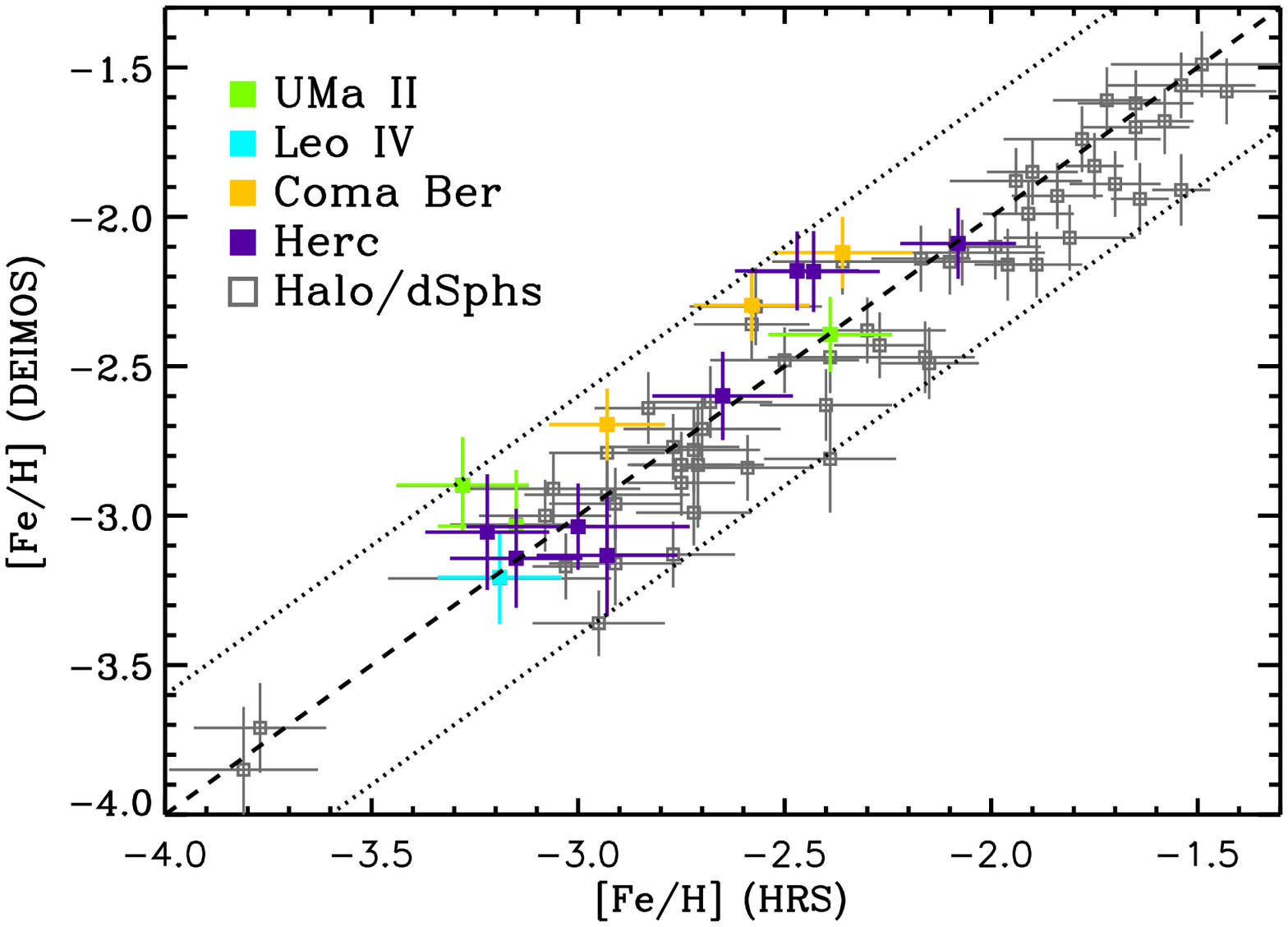}	
\label{subfig: mrshrsfeh}}	
\qquad
\subfloat{
\includegraphics[width=.47\textwidth]{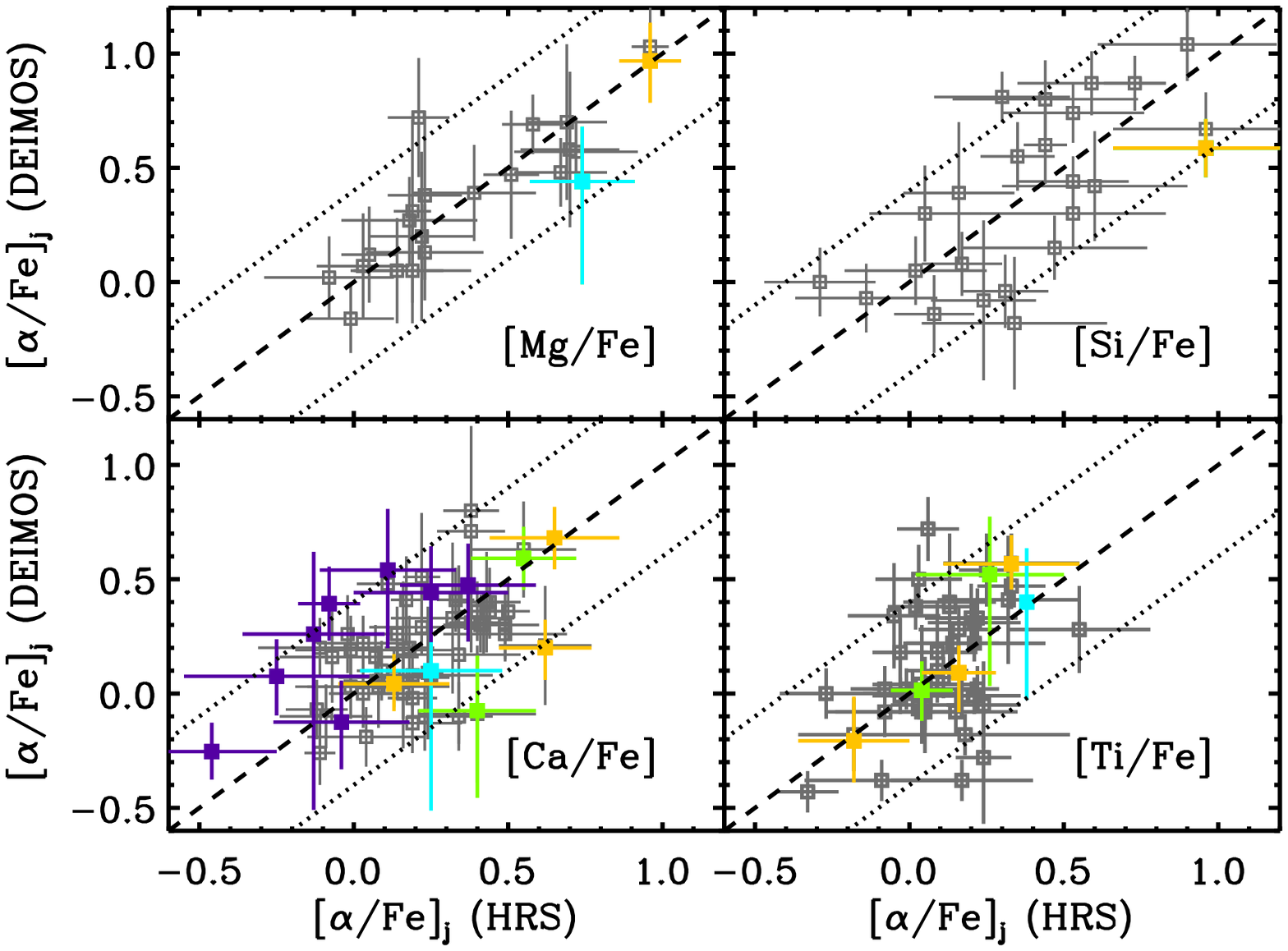}	
\label{subfig: mrshrsalphafe}}	
\caption{\textit{Left:}  
We compare our DEIMOS [Fe/H] measurements 
against published high$-$resolution (HRS) measurements 
in Com~Ber, UMa~II \citep{Frebel10b}, Leo~IV \citep{Simon10}, 
and Herc \citep{Aden11}. We add to the comparison 
halo and classical dSphs stars presented in \citet{Kirby10}.
\textit{Right:} Comparison of [Mg/Fe], [Si/Fe], 
[Ca/Fe], and [Ti/Fe] abundance ratios for the same sample. 
The \xfe\, abundance uncertainties for the UFD stars were 
measured as described in \S\,\ref{abunds_ssec}. 
All measurements have been transformed to the \citet{Asplund09} 
abundance scale. The dotted lines indicate a difference of 
$+0.4$ and $-0.4$ from equality, comparable to our largest 
allowed uncertainty. Both panels show good agreement 
between DEIMOS and HRS measurements.
(see \S\,\ref{abunds_ssec}). }
\label{fig: mrshrs}
\end{figure*}

In Figure \ref{fig: alphaatm}, we compare [$\alpha$/Fe]$_{\rm{atm}}$ 
against the weighted mean of \xfe\, for the nine stars with measurements 
for all elements. The comparison shows that [$\alpha$/Fe]$_{\rm{atm}}$ 
is offset relative to the weighted mean by $-0.063\pm{0.010}$ dex.
We attribute this offset to the influence of the [Mg/Fe] measurements, 
which are systematically higher than [$\alpha$/Fe]$_{\rm{atm}}$ for all stars 
in this subsample. While the mean assigns equal weight to each element 
when the uncertainties are comparable, the measurement of 
[$\alpha$/Fe]$_{\rm{atm}}$ is less affected by the Mg abundance 
due to the relatively small number of Mg lines in the DEIMOS spectrum. 
We adjust the definition of \alphafeavg\, as \mbox{[$\alpha$/Fe]$_{\rm{atm}}$ + 
0.063 dex} in order to account for the systematic discrepancy described 
above. 

We note that although use of an \alphafeavg\, 
blurs nuanced differences that may be present between the 
different elements, it is a useful quantity because of the closely
related nucleosynthetic origin of these elements. 
We report \alphafeavg\, measurements for \nstars\ stars (equal to
the number of \alphafeatm\, measurements) including 
seven stars for which no individual \xfe\, were detect due to 
a lack of signal. Table \ref{table: ufd_data} summarizes basic 
properties for each UFD, the number of stars with available 
\alphafe\, measurements, and the weighed average metallicity 
for each UFD using only these stars.

\subsection{Comparison with High$-$Resolution Studies}\label{compare_ssec}

To validate our technique, we compare our results 
against high$-$resolution (HRS) abundances for overlapping 
stars in Com~Ber, UMa~II \citep{Frebel10b}, Leo~IV \citep{Simon10}, 
and Herc \citep{Aden11}. 
Figure \ref{fig: mrshrs} shows the results of this comparison for [Fe/H] 
(left panel) and the individual \xfe\, abundance ratios (right panel), where
all abundances have been placed in the \citet{Asplund09} abundance
scale. Due to the small number of matching stars, we add to the comparison 
a set of halo stars with DEIMOS and high$-$resolution measurements 
analyzed by \citet{Kirby10} using the same technique. We note that our 
modification to \citet{Kirby10}'s approach lies in the determination 
of abundance \textit{uncertainties}, and hence does not affect the 
comparison in Figure \ref{fig: mrshrs}.

We find good agreement between [Fe/H] in both samples, 
with a mean difference of $-0.035\pm{0.022}$ dex in the sense 
\mbox{HRS$-$DEIMOS}, where the uncertainty is the standard 
error of the mean. The individual abundance 
ratios do not show any systematic offsets. The mean differences for 
[Mg/Fe], [Si/Fe], [Ca/Fe] and [Ti/Fe] are \mbox{$+0.018\pm{0.035}$}, 
\mbox{$-0.005\pm{0.057}$}, \mbox{$-0.004\pm{0.029}$}, and 
\mbox{$0.022\pm{0.033}$} dex, respectively, demonstrating that 
we obtain accurate abundances over our entire range of values.

\begin{deluxetable*}{lcrcccccccr}[t!]
\tablecolumns{11}
\tabletypesize{\footnotesize}
\tablecaption{UFD Basic Data}
\tablewidth{0pt}
\tablehead{
\colhead{UFD} & 
\colhead{R$_{\rm{helio}}^{a}$} (kpc) & 
\colhead{V$_{\rm{rad}}^{b}$ (km/s)} & 
\colhead{$\sigma_{\rm{V}}^{b}$ (km/s)} & 
\colhead{M$_{{V}}{}^{c}$} & 
\colhead{N$_{\rm{Mg}}$} & 
\colhead{N$_{\rm{Si}}$} & 
\colhead{N$_{\rm{Ca}}$} & 
\colhead{N$_{\rm{Ti}}$} &
\colhead{N$_{\alphafeavg}$} & 
\colhead{$\langle\rm{[Fe/H]}\rangle^{d}$}
}
\startdata
Segue 1 &  23 & 208 $\pm$ 0.9 & 3.7 $\pm$ 1.4 & $-$1.5 & 3 & 4 & 5 & 4 & 5 & $-2.03\pm0.06$ \\
Coma Berenices &  44 & 98 $\pm$ 0.9 & 4.6 $\pm$ 0.8 & $-$4.1 & 1 & 5 & 7 & 4 & 9 & $-2.53\pm0.06$ \\
Ursa Major II &  32 & $-$116 $\pm$ 1.9 & 6.7 $\pm$ 1.4 & $-$4.2 & 2 & 5 & 4 & 4 & 6 & $-2.15\pm0.06$ \\
Canes Venatici II & 151 & $-$128 $\pm$ 1.2 & 4.6 $\pm$ 1.0 & $-$4.9 & 1 & 2 & 4 & 5 & 8 & $-2.18\pm0.06$ \\
Leo IV & 158 & 132 $\pm$ 1.4 & 3.3 $\pm$ 1.7 & $-$5.0 & 0 & 4 & 2 & 4 & 4 & $-2.89\pm0.11$ \\
Ursa Major I & 106 & $-$55 $\pm$ 1.4 & 7.6 $\pm$ 1.0 & $-$5.5 & 2 & 7 & 9 & 8 & 11 & $-2.04\pm0.05$ \\
Hercules & 138 & 45 $\pm$ 1.1 & 5.1 $\pm$ 0.9 & $-$6.6 & 1 & 5 & 13 & 3 & 13 & $-2.42\pm0.05$ \\
Leo T & 417 & 38 $\pm$ 2.0 & 7.5 $\pm$ 1.6 & $-$7.1 & 0 & 2 & 1 & 4 & 5 & $-1.94\pm0.08$
\enddata
\renewcommand{\thefootnote}{\alph{footnote}}
\footnotetext[1]{R$_{\rm{helio}}$ from \citet{Martin08b}; see references therein for each UFD.}
\footnotetext[2]{V$_{\rm{rad}}$ and $\sigma_{\rm{V}}$ taken from \citet{Simon07} except for Segue1, taken from \citet{Simon11}.}
\footnotetext[3]{$M_{\rm{V}}$ from \citet{Martin08b} except for Leo T, taken from \citet{Irwin07}.}
\footnotetext[4]{$\langle\rm{[Fe/H]}\rangle$ is the mean metallicity for each UFD using only the stars with good \alphafeavg\, measurements (see \S\,\ref{abunds_ssec}.)}
\label{table: ufd_data}
\renewcommand{\thefootnote}{\arabic{footnote}}
\end{deluxetable*}

\section{Abundance Results I: Individual UFDs}\label{results_sec_individual}

\begin{figure*}[t!]
\centering
\includegraphics[width=\textwidth]{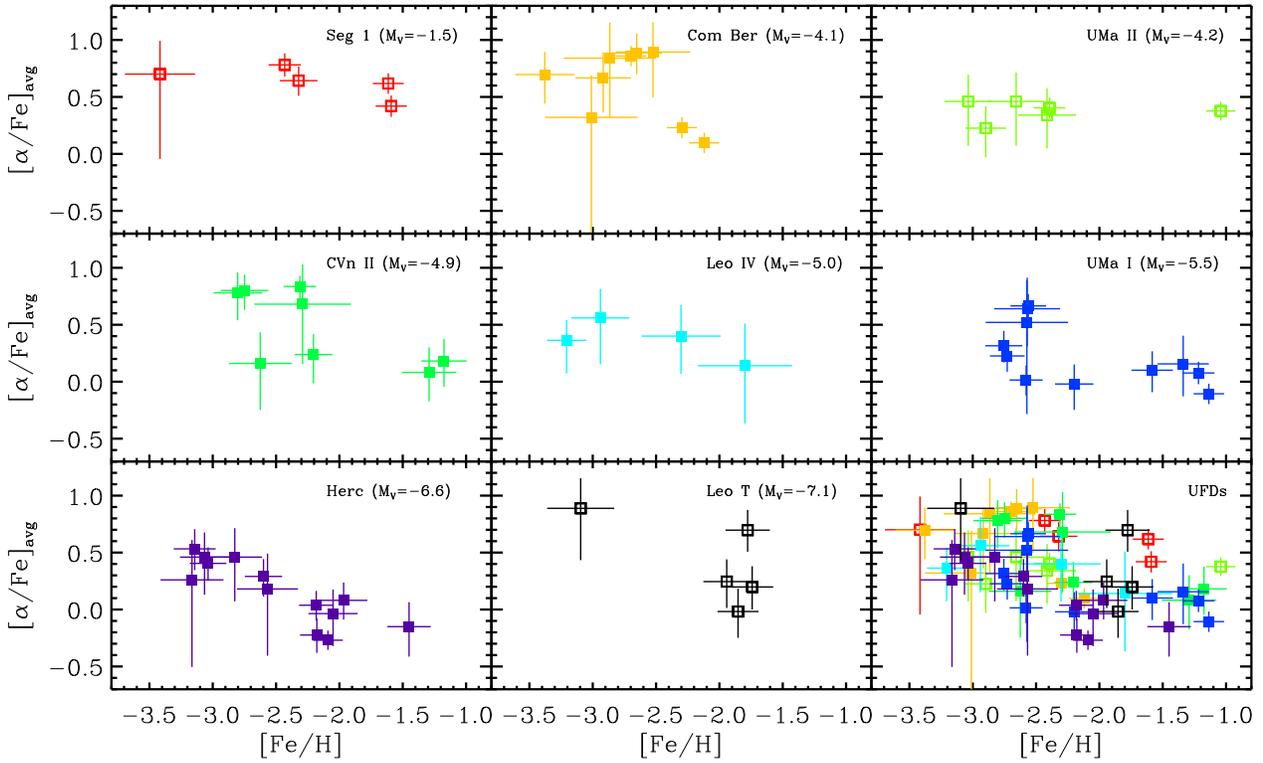}
\caption{\alphafeavg\, as a function of [Fe/H] for each UFD in our sample,
ordered by increasing galaxy luminosity. The bottom$-$right panel shows
the combined UFD sample, also seen in Figure \ref{fig: alphafe_ufds_all}.
Each UFD is assigned a different color. We use the same color scheme
throughout the paper. Five of the UFDs have ancient stellar populations
\textit{and} show an overall decrease in \alphafeavg\, with increasing [Fe/H]. 
We discuss their evolution in \S\,\ref{sssec_decreasing}. We use 
filled squares for stars in these galaxies and empty squares for the 
others. The presence of low \alphafeavg\, stars at higher [Fe/H] strongly 
suggests that Type~Ia SNe chemically enriched the ISM of these UFDs. 
Due to the time delay for Type~Ia SNe after the onset of star formation, 
$t_{\rm{min,Ia}}$, our data suggests that most UFDs underwent at least 
a limited time period of extended chemical evolution, no shorter than 
$t_{\rm{min,Ia}}\sim\tsnemin$ Myr.}
\label{fig: alphafe_ufds}
\end{figure*}

The alpha abundances reflect the enrichment from SNe, 
and thus help constrain the underlying star formation 
history of a galaxy. In this section, we highlight the most 
salient qualitative trends for each UFD. We present the abundance 
measurements for all eight UFDs in Table~\ref{table: abundances}. 
Figure~\ref{fig: alphafe_ufds} shows 
the \alphafeavg$-$[Fe/H] trends for each UFD in our sample, 
in order of increasing luminosity. The stars in each of our UFDs 
spans a range in metallicity greater than 1~dex. We discuss 
the implications of these trends in \S\,\ref{disc_indiv}.  

\textbf{Segue~1.$-$} Seg~1 is the faintest and nearest UFD 
known to date. We measure \alphafe\, abundances for the 5 
stars for which S11 reports metallicities (excluding the star 
with only an upper metallicity limit, \mbox{[Fe/H] $< -3.4$}). 
In spite of its low luminosity, the spread in metallicity is remarkably 
large, spanning $\gtrsim{2}$ dex from our 5 stars alone. 
The \alphafeavg\, abundance ratios do not show any noticeable 
decrease with [Fe/H] and are roughly constant at $\sim{+0.6}$ dex; 
a similar trend is seen in [Ca/Fe]. The two stars at $\rm{[Fe/H]}\sim{-1.6}$ 
have slightly lower [Si/Fe] and [Ti/Fe] abundances than the two stars at $\rm{[Fe/H]}\sim{-2.4}$ 
(only by $\sim{0.15}$). In summary, Seg~1 shows enhanced abundance ratios 
even up to \mbox{[Fe/H]$\sim{-1.6}$}, suggestive of a lack of pollution by Type~Ia 
SNe. The only published \alphafe\, abundance ratios in Seg~1 are those 
of \citet{Norris10c}. Using high$-$resolution, 
they report \mbox{[Mg/Fe] = +0.94}, \mbox{[Si/Fe] = +0.80}, 
\mbox{[Ca/Fe] = + 0.84}, and \mbox{[Ti II/Fe] = +0.65} for a
\mbox{[Fe/H] = $-$3.52} CEMP-no\footnote{CEMP-no: Carbon$-$enhanced, 
metal$-$poor star without heavy neutron element enhancements, 
see summary of CEMP nomenclature in \citet{Norris13}} star (not in our sample). 
Their measurement agrees with the abundances measured in our most 
metal$-$poor star, which has a comparable metallicity\footnote{
We cannot comment on the CEMP classification of our star, but
note that a significant fraction of metal$-$poor stars are 
carbon$-$enhanced \citep{Norris13}. Thus, we caution the reader
that this abundance comparison may only be fully warranted if our
star is also a CEMP$-$no object.}.

\textbf{Coma Berenices.$-$} Although there is no published 
constraint on its age spread, Com~Ber's CMD appears consistent 
with a very old age, with no intermediate age stars (Figure 3 of \citealt{Munoz10}). 
Com~Ber shows high \alphafeavg\,, greater than $ +0.4$, at 
lower [Fe/H]  and lower \alphafeavg\, by $\sim{0.4}$ dex for the two highest
[Fe/H] stars. [Si/Fe] and [Ca/Fe] also appear higher towards lower [Fe/H]. 
We do not detect any clear trend in  [Ti/Fe], and there is insufficient data for [Mg/Fe]. Using 
high$-$resolution spectroscopy, \citet{Frebel10b} also reported enhanced alpha 
abundances at \mbox{[Fe/H] $<-2.5$}, whereas their most metal$-$rich star shows 
systematic lower \alphafe\, abundance ratios by $\sim{0.4}$ dex. Their results show 
broad agreement with ours. 

\textbf{Ursa~Major~II.$-$}  As for Com~Ber, the CMD of UMa~II is 
suggestive of a very old stellar population with no intermediate age 
stars \citep{Munoz10}. In contrast to Com~Ber, UMa~II shows signs 
of tidal stripping, suggesting it may have originally been a more luminous 
satellite. All of our \alphafeavg\, measurements cluster at \alphafeavg$\sim{+0.4}$, 
spanning a large range of metallicities up to $\rm{[Fe/H]}\sim{-1.1}$. 
The three most metal$-$poor stars are overabundant in [Si/Fe] by $\sim{0.3}$ 
relative to the rest of the sample. Except for [Si/Fe], other abundance 
ratios appear to have flat abundance patterns. \citet{Frebel10b}'s 
measurements of three stars are in agreement with our result. 
They show roughly constant abundance ratios for [Mg/Fe], [Ca/Fe], 
and [Ti/Fe] for their three stars, all with \mbox{[Fe/H] $< -2.3$}. We 
measure \mbox{$\rm{[Si/Ca]} = +1.17\pm{0.37}$} for a single star, 
which was also studied by \citet{Frebel10b} (their UMa$-$S2). They only 
measure an upper limit on [Si/Fe], \mbox{$\rm{[Si/Fe]} < +1.46$}. 
In combination with their [Ca/Fe] measurement, their upper limit
for [Si/Ca] is $+1.08$, in agreement with our measurement. We 
defer the discussion of anomalous abundance ratios to \S\,\ref{ssec_anomalous}.

\textbf{Canes~Venatici~II.$-$} The next five UFDs are at considerably
larger distances than the previous three (see Table \ref{table: ufd_data}). 
Ground$-$based photometry suggests that CVn~II is composed exclusively 
of an old ($>$ 10 Gy) stellar population \citep{Sand12,Okamoto12}. We 
present for the first time \alphafe\, abundance ratios for this galaxy. At 
$\rm{[Fe/H]} < -2$, we find both high and low \alphafeavg\, abundance ratios,
hinting at some intrinsic scatter. On average, \alphafeavg\, is higher at lower
[Fe/H]. The distribution of [Ca/Fe] and [Ti/Fe] abundance ratios tentatively 
supports the presence of significant scatter at low [Fe/H]. 

\textbf{Leo~IV.$-$} \citet{Brown12} have recently constrained the 
spread of ages of the stellar population to less than $\sim{2}$~Gyr. 
We have measured \alphafeavg\, for four stars, which are consistent with 
either a shallow increase in \alphafeavg\, with decreasing [Fe/H], or
a constant enhancement of $\sim{0.3}$~dex. [Si/Fe] shows some evidence for 
slightly higher abundance ratios, $> 0.5$~dex, but again no trend with 
[Fe/H] can be discerned. [Ti/Fe] is likewise relatively high. [Ca/Fe], measured 
in only two stars, is comparable to \alphafeavg. We note that the 
larger uncertainties in all abundance ratios (relative to other UFDs) are 
due to the low S/N of the DEIMOS spectra for this satellite. In agreement with 
our result, \citet{Simon10} report enhanced \alphafe\, abundance ratios for 
the brightest RGB, Leo IV~S1. This star is also included in our sample; 
a comparison of the abundance ratios can be seen in Figure \ref{fig: mrshrs}.

\begin{deluxetable*}{lrrrrrrrr}
\tablecolumns{9}
\tabletypesize{\footnotesize}
\tablecaption{Abundance Results}
\tablehead{
\colhead{UFD} & 
\colhead{RA (J2000)} & 
\colhead{DEC (J2000)} & 
\colhead{[Fe/H]} & 
\colhead{\alphafeavg} & 
\colhead{[Mg/Fe]} & 
\colhead{[Si/Fe]} & 
\colhead{[Ca/Fe]} & 
\colhead{[Ti/Fe]}
}
\startdata
Seg 1 & $10:06:52.33$ & $+16:02:35.8$ & $-$3.42$\pm$0.28 & +0.70$^{+0.29}_{-0.74}$ &  &  & +0.76$^{+0.29}_{-0.53}$ &  \\[+2pt]
Seg 1 & $10:07:10.08$ & $+16:06:23.9$ & $-$1.61$\pm$0.12 & +0.62$^{+0.09}_{-0.09}$ & +0.70$^{+0.12}_{-0.14}$ & +0.68$^{+0.12}_{-0.14}$ & +0.58$^{+0.13}_{-0.13}$ & +0.39$^{+0.11}_{-0.13}$ \\[+2pt]
Seg 1 & $10:07:14.58$ & $+16:01:54.5$ & $-$1.59$\pm$0.12 & +0.42$^{+0.09}_{-0.09}$ & +0.43$^{+0.14}_{-0.14}$ & +0.52$^{+0.12}_{-0.11}$ & +0.33$^{+0.14}_{-0.17}$ & +0.25$^{+0.12}_{-0.13}$ \\[+2pt]
Seg 1 & $10:07:42.71$ & $+16:01:06.9$ & $-$2.43$\pm$0.13 & +0.78$^{+0.10}_{-0.10}$ & +0.87$^{+0.20}_{-0.31}$ & +0.99$^{+0.14}_{-0.13}$ & +0.59$^{+0.14}_{-0.14}$ & +0.71$^{+0.15}_{-0.19}$ \\[+2pt]
Seg 1 & $10:07:02.46$ & $+15:50:55.2$ & $-$2.32$\pm$0.15 & +0.64$^{+0.12}_{-0.13}$ &  & +0.74$^{+0.17}_{-0.25}$ & +0.66$^{+0.19}_{-0.19}$ & +0.58$^{+0.20}_{-0.29}$ \\[+2pt]
Com Ber & $12:26:29.01$ & $+24:04:03.8$ & $-$2.52$\pm$0.29 & +0.89$^{+0.26}_{-0.40}$ &  &  &  &  \\[+2pt]
Com Ber & $12:26:45.14$ & $+23:50:44.7$ & $-$2.92$\pm$0.22 & +0.67$^{+0.23}_{-0.30}$ &  & +1.07$^{+0.24}_{-0.33}$ & +0.26$^{+0.34}_{-0.71}$ & +0.76$^{+0.31}_{-0.69}$ \\[+2pt]
Com Ber & $12:26:55.46$ & $+23:56:09.8$ & $-$2.70$\pm$0.12 & +0.86$^{+0.09}_{-0.09}$ & +0.97$^{+0.17}_{-0.18}$ & +1.20$^{+0.11}_{-0.12}$ & +0.68$^{+0.14}_{-0.14}$ & +0.57$^{+0.13}_{-0.11}$ \\[+2pt]
... & ... & ... & ... & ... & ... & ...
\enddata
\label{table: abundances}
\tablecomments{Table~\ref{table: abundances} is published in its entirety in the electronic edition of the Astrophysical Journal. A portion is shown here for
guidance regarding its form and content.}
\end{deluxetable*}

\textbf{Ursa~Major~I.$-$} \citet{Brown12} 
have shown that the stellar population is ancient, 
and constrained the spread in ages to less than 
$\sim{2}$ Gyr. We present the first \alphafe\, abundance ratios 
measured in UMa~I. The \alphafeavg\, abundance 
pattern for UMa~I shows \textit{on average} increasing 
\alphafeavg\, abundance ratios towards lower [Fe/H], with 
the possible exception of [Ca/Fe]. There is a hint of increased 
intrinsic scatter in \alphafeavg\, and [Ca/Fe] at low [Fe/H], 
indicating that this galaxy might have experienced inhomogeneous 
chemical enrichment.

\textbf{Hercules.$-$} \citet{Brown12} have constrained 
the age and age spread in star formation to be similar 
to that in Leo~IV and UMa~I. We present measurements 
of \alphafeavg\, for 13 stars, currently the largest published 
sample of \alphafe\, for this UFD. One of our stars has 
\mbox{$\rm{[Mg/Ca]} = +0.72\pm{0.21}$}. We discuss its 
abundance pattern further in \S\,\ref{ssec_anomalous}.
Herc shows a clear trend for rising \alphafeavg, [Si/Fe], 
and [Ca/Fe] towards lower [Fe/H], with little 
scatter, reaching \alphafeavg$\sim{+0.5}$ at the lowest [Fe/H]. The 
\alphafeavg\, enhancement seems systematically lower at fixed [Fe/H] 
than in Seg~1 and Com~Ber. The data is insufficient to suggest any pattern 
in the case of [Mg/Fe] and [Ti/Fe]. Recently, \citet{Aden11} reported high$-$resolution 
[Ca/Fe] abundance ratios for 10 RGB stars in Hercules (eight overlap with our 
sample) with [Ca/Fe] varying from $\sim{+0.3}$ at \mbox{[Fe/H]$\sim{-3.1}$} to 
\mbox{[Ca/Fe]$\sim{-0.5}$} at \mbox{[Fe/H] $\sim{-2}$}, concluding that
Herc experienced very inefficient star formation. Our measurements 
confirm the trend of decreasing [Ca/Fe] with rising [Fe/H].

\textbf{Leo~T}.$-$ Leo~T \citep{Irwin07} is the only UFD 
with evidence for recent star formation 
\citep[e.g.,][]{Weisz12,Clementini12}. It also has a 
large amount of HI gas \citep{Ryan-Weber08}. These two properties distinguish it
from all the other UFDs in this study. We have measured \alphafeavg\, for 5 stars, 
4 of which cluster around \mbox{[Fe/H]$\sim{-2}$}. and have a range of \alphafeavg\, 
from $\sim{-0.1}$ to $\sim{+0.7}$. The only element with $> 2$ measurements 
is [Ti/Fe], which was measured for the 4 stars at \mbox{[Fe/H]$\sim{-2}$}. All [Ti/Fe] 
measurements cluster between $-0.1$ and $+0.3$ dex. The presence
of low \alphafe\, stars is expected for systems with extended star formation.

\textbf{In summary}, we observe the following trends 
for the individual UFDs:

\begin{itemize}[noitemsep]
\renewcommand{\labelitemi}{$\bullet$}
\item All UFDs have on average 
high \alphafe\, abundance ratios ($\gtrsim{+0.3}$) at [Fe/H] $< -2.5$. 
High \alphafe\, abundance ratios are consistent with chemical enrichment 
by Type~II SNe.
\item Most stars with [Fe/H] $> -1.5$ (excluding our Seg~1 and UMa~II samples)
have relatively low \alphafeavg\, abundance ratios, \alphafeavg $< +0.4$, 
suggesting that chemical evolution lasted at least as long as the 
minimum time delay for Type~Ia SNe. 
\item Seg~1 and UMa~II are alpha$-$enhanced across their entire 
metallicity range. They do not show a statistically significant decrease 
in \alphafe\, abundance ratios as a function of [Fe/H], in contrast to the 
other UFDs. 
\item The degree of alpha enhancement shows some hint of being 
different between UFDs, with Com~Ber and Seg~1 having higher \alphafe\, 
abundance ratios than Herc or UMa~II at $\rm{[Fe/H]}\sim{-3}$. This could 
be a reflection of a different mix of SNe across the various UFDs, stochastic 
sampling of the same IMF, and/or inhomogeneous mixing. 

\end{itemize}

\section{Abundance Results II: The Ensemble of UFDs}\label{results_sec_all}

A comparison of chemical abundances 
can shed light on the relationship between different stellar 
populations. It has been previously shown that the 
\alphafe\,$-$[Fe/H] pattern in the classical dSphs disagrees 
with the Milky Way inner halo pattern for \mbox{[Fe/H]$\gtrsim{-2}$}. 
Building on this difference, simulations 
by \citet{Robertson05} have suggested that the major building 
blocks of the inner halo had different star formation history 
than the extant classical dwarf galaxies.

Here, we compare the abundance patterns in our UFD 
sample against the inner halo, and also against one 
classical dSph, CVn~I. With $\mathrm{M}_{\mathrm{V}} = -8.6$ \citep{Martin08b}, 
CVn~I is $\sim{2}$ magnitudes brighter than Leo~T (the brightest UFD 
in our sample) and it is composed primarily of an old \mbox{($\gtrsim{10}$ Gy)} 
population \citep{Martin08a,Okamoto12}. We use the 
CVn~I sample from \citet{Kirby10}, reanalyzed to reflect our
updated \textit{uncertainty} analysis, which does not assume 
symmetric uncertainties (\S\,\ref{abunds_ssec}). We 
include the reanalyzed abundance measurements 
for CVn~I (referenced to the \citealt{Asplund09} 
solar abundance scale) at the bottom of Table~\ref{table: abundances}.
The CVn~I sample actually extends to $\rm{[Fe/H]}<-3$ and can 
be used as a comparison sample to the UFDs. For the halo, 
we rely on the chemical abundance compilation by \citet{Frebel10d}.

Since most of the UFDs have similar observed 
abundance trends, we merge the samples for 
the different UFDs to obtain a combined 
sample of more than 30 stars for each element, 
excluding Mg (due to very weak lines, Mg is only 
detectable in \nmg\, stars). Figure \ref{fig: alphafe_ufds_all} 
(left panels) compares the \alphafe\, abundance ratios 
for our combined UFD sample against the inner halo 
population. For the halo sample, we calculated 
\alphafeavg\, as the mean of the available [Mg/Fe], [Si/Fe], 
[Ca/Fe], and [Ti/Fe] abundance ratios. The right panels show
a comparison of the \alphafe\, abundance patterns of the more
massive dSph CVn~I against Milky Way inner halo stars. We explain 
our statistical comparison method in \S\,\ref{mcmc_ssec}, and 
describe the results in \S\,\ref{mcmc_results}. We
comment on the presence of two stars with anomalous
abundance ratios in \S\,\ref{ssec_anomalous}.  

\subsection{MCMC Modeling of Empirical \xfe$-$$\rm{[Fe/H]}$ Trends}\label{mcmc_ssec}

In order to identify the best$-$fitting trend in \alphafe\,$-$[Fe/H] space in a 
statistically robust way, we define simple parameterizations of the various 
trends predicted by chemical evolution models. In these models, 
the ISM is quickly enriched by ejecta from Type II SNe, resulting in 
high \alphafe\, at low [Fe/H]. The onset of Type Ia SNe ejects more Fe$-$peak 
rich material and acts to lower \alphafe. Due to the delayed onset of Type Ia 
relative to Type II SNe, the change in \alphafe\, can be seen as a turnover or 
knee at a particular [Fe/H]. Afterwards, the decrease in \alphafe\, is modulated by
the number of Type II and Type Ia SNe that explode. It is possible to increase
\alphafe\, with a late$-$time starburst \citep{Gilmore91}.

\subsubsection{Empirical \xfe$-$$\rm{[Fe/H]}$ Models}

We consider three simple models describing a path in 
$\xfe\,-\rm{[Fe/H]}$ space, where [Fe/H] and \xfe\, are 
defined as the $x$ and $y$ coordinates. The "Constant \alphafe\, Model" 
(Model A) is a single$-$parameter model with a constant value of \xfe\, at 
all [Fe/H], i.e. a flat line with $y=y_{0}$. It is representative of Type II SNe 
enrichment. The "Single Slope Model" (Model B) is a two parameter linear model with 
freely$-$adjustable slope, $m = dy/dx$ and y$-$intercept, $b$, $y(x) = m\,x+b$. 
This model is representative of Type Ia SNe enrichment. 
 
Equations~\ref{eqn_A1}$-$\ref{eqn_A2} parameterize 
the "Knee Model" (Model C),  which is a combination of 
flat and decreasing \xfe\, segments. It represents
early Type~II SNe enrichment followed by a phase
where Type~Ia SNe contributed to the chemical evolution.

\begin{align}
y (x) = m\,{\rm{[Fe/H]_{0}}} + b\,;\, x \leq \rm{[Fe/H]_{0}}\label{eqn_A1}\\
y (x) = m\,x + b\,;\, x > \rm{[Fe/H]_{0}}\label{eqn_A2}
\end{align}

Here, $\rm{[Fe/H]}_{0}$ defines the boundary between the two
segments, and is typically referred to as the knee. Parameters 
$m$ and $b$ are defined as in Model B.

\begin{figure*}[t!]
\centering
\includegraphics[width=0.9\textwidth]{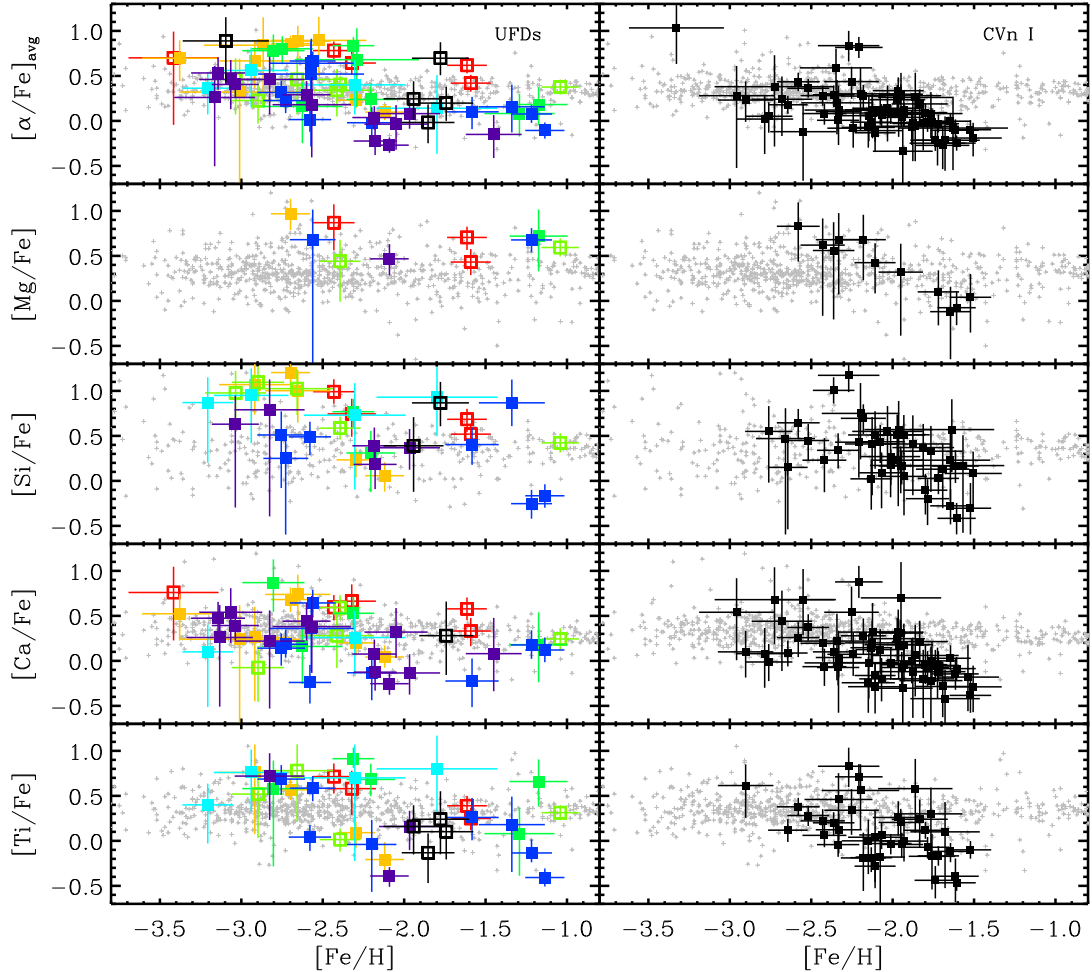}
\caption{We compare \alphafeavg\, abundance ratios in 
the UFDs (\textit{left panels}: filled/empty colored squares, 
same color and point scheme as in Figure~\ref{fig: alphafe_ufds}), 
the Milky Way inner halo (small gray crosses \textit{in both panels}), and the 
CVn~I classical dSph (\textit{right panels}: filled black squares). 
In the left panels, we ask whether the UFD population 
shares a similar abundance pattern as the inner halo. In 
the right panels, we compare CVn~I and the inner halo. 
From top to bottom, we plot the \alphafeavg, [Mg/Fe], [Si/Fe], 
[Ca/Fe] \& [Ti/Fe] abundance ratios. For the inner halo, 
we rely on the metal$-$poor star abundance 
compilation by \citet{Frebel10d}, with all abundances referenced 
to the \citet{Asplund09} solar abundance scale. For CVn~I, we 
use \citealt{Kirby10}'s~CVn~I sample, reanalyzed using our 
updated analysis technique (\S\,\ref{analysis_sec}). We 
qualitatively summarize the results of the statistical comparison
described in \S\,\ref{results_sec_all}. Both the UFDs and 
CVn~I show lower \alphafe\, abundances than
the inner halo at higher [Fe/H]. In addition, CVn~I shows a
hint of a turnover in \alphafe\, at a metallicity between $-$2.5 
and $-$2.0, not detected for the UFD sample (regardless
of using the entire UFD sample, or the subsample marked with
filled squares). This provides 
a hint that the (less luminous) UFDs underwent less efficient 
star formation than the classical dSphs, based on
the non$-$detection of a turnover at \mbox{$\rm{[Fe/H]} > -3$}.
Our data highlights that both UFDs and classical dSphs
have a different chemical evolution than the inner halo.} 
\label{fig: alphafe_ufds_all}
\end{figure*}

\subsubsection{Probability Distribution Functions for each Model}
We seek to calculate the best$-$fit parameters and associated
confidence intervals for each model given a dataset $\mathcal{D}$, 
where $\mathcal{D}$ is a set of [Fe/H] and \xfe\,abundances. 
\textit{Due to our asymmetric uncertainties, we cannot rely on a simple regression 
analysis}. For this purpose, we use a Markov$-$chain Monte Carlo method
to calculate P\,($\boldsymbol\theta$|$\mathcal{D}$), the joint probability density 
functions for each model as a function of its parameters $\boldsymbol\theta$
given $\mathcal{D}$. Specifically, we measure the probability density 
functions ${\rm{P}}_{\rm{A}}(y_{0})$, ${\rm{P}}_{\rm{B}}\,(m,k_{\rm{icpt}}$), and 
${\rm{P}}_{\rm{C}}\,(m,k_{\rm{icpt}},\rm{[Fe/H]_{0}})$ for models A, B, and C.
The primary input to the Markov chain are likelihoods for the full dataset $\mathcal{D}$ given 
a realization of $\boldsymbol\theta$, $\mathcal{L}(\mathcal{D}|\boldsymbol\theta$). 
This in turn requires calculating \like$_{i}$, the likelihood of star $i$ being 
drawn from the model. 

Due to the non$-$Gaussian \alphafeatm\, and 
\xfe\, uncertainties discussed in \S\ref{abunds_ssec}, 
we make use of the probability distribution for each 
abundance to calculate \like$_{i}$. We denote these 
probability distributions as F, to avoid confusion with 
P\,($\boldsymbol\theta$|$\mathcal{D}$). 
We compute the random component of F\,(\xfe) from the 
$\chi^{2}$ contours described in \S\,\ref{abunds_ssec}. 
The probability of a star having abundance $y$ 
is given by Equation~\ref{eqn: pdf_ran}, and peaks at 
$\chi^{2}_{\rm{min}}$.

\begin{equation}
{\rm{F_{ran}}}(y) \propto{} \exp{\left[-\frac{1}{2}\left(\chi^{2}(y)-\chi^{2}_{\rm{min}}\right)\right]}
\label{eqn: pdf_ran}
\end{equation}

We incorporate the systematic uncertainty in 
each measurement by convolving F$_{\rm{ran}}$ 
with a Gaussian with zero mean and standard 
deviation equal to $\sigma_{\rm{sys}}$. The full probability
function for \xfe\, is given in Equation \ref{eqn: pdf_tot}.

\begin{equation}
{\rm{F}}\,(y) = \displaystyle\int{{\rm{F_{ran}}}\,(y')\frac{1}{\sqrt{2\pi\sigma^{2}_{\rm{sys}}}}{\rm{exp}}\left[-\frac{1}{2}\left(\frac{y-y'}{\sigma_{sys}}\right)^{2}\right]\,dy'}
\label{eqn: pdf_tot}
\end{equation}

In contrast to \xfe\,, the [Fe/H] $\chi^{2}$ contours 
are symmetric for total uncertainties up to 
$\sigma_{\rm{[Fe/H]}} \sim{0.4}$ (\S\,\ref{tefffeh_ssec}). 
We thus define the probability distribution for 
[Fe/H], ${\rm{F}}\,(x)$, as a Gaussian with \mbox{$\sigma_{\rm{[Fe/H]}}$}, 
centered on the best$-$fit [Fe/H] value. We now calculate 
\like$_{i}$ using Equation~\ref{eqn: integral_pdf}. 

\begin{equation}
\like_{i}\,\propto \displaystyle\int{{\rm{F}}\,(x)\,{\rm{F}}\,(y\,(x))\,dx}
\label{eqn: integral_pdf}
\end{equation}

The full likelihood $\mathcal{L}(\mathcal{D}|\boldsymbol\theta$) is
the product of the individual likelihoods for stars 
\mbox{1 $\leq$ $i$ $\leq$ N},  $\like =  \prod^{N}_{i=1} \like_{i}$. 
We run the Markov chain using the Metropolis-Hastings algorithm
with a Gaussian$-$distributed kernel.
We constrain [Fe/H]$_{0}$ to lie more than 0.3 dex away 
from the minimum in the sample, and below $\rm{[Fe/H]} = -2$. 
We chose this prior by noting that no knee has been observed 
in brighter dSphs at higher metallicities.

We constrain the slope in Models B and C 
to $m< 0$. We run each chain for 100,000 steps for models 
A and B, and 250,000 steps for Model C.  The larger number of 
steps is needed to better sample the larger parameter space
in this model. In all cases, the first 1,000 steps are discarded as a 
burn$-$in period. For each step $k$, we compute the ratio of 
likelihoods $r\equiv\mathcal{L}_{k}/\mathcal{L}_{k-1}$ 
between the $k$ and $k-1$ steps\footnote{The ratio $r$ is actually 
defined using the ratio of posterior probabilities for parameters 
$\boldsymbol\theta_{k}$ and $\boldsymbol\theta_{k-1}$ given data $\mathcal{D}$,
$\rm{P}(\boldsymbol\theta_{k}|\mathcal{D})/\rm{P}(\boldsymbol\theta_{k-1}|\mathcal{D})$. 
P and $\mathcal{L}$ are related by Bayes' Theorem as 
${\rm{P}}_{k}(\theta|\mathcal{D}),\propto\,\mathcal{L}_{k}\,
(\mathcal{D}|\boldsymbol\theta)\,\pi(\boldsymbol\theta_{k})$, 
where $\pi(\boldsymbol\theta_{k})$ is the prior probability of the set of
parameters $\boldsymbol\theta_{k}$. Under our assumption of 
uniform priors, the two ratios are identical.}

We accept the new set of parameters if $r > 1$ or \mbox{$0< \mathcal{U}\,(0,1) < r < 1$}, 
where $\mathcal{U}\,(0,1)$ is a uniform deviate between 0 and 1. 
Otherwise, we reject the trial step and save the parameters from 
step $k-1$ in step $k$.

The density of points in the chain defines P($\boldsymbol\theta|\mathcal{D}$). 
We determine the best$-$fit parameters from the peak of the 
one$-$dimensional probability distribution of \textit{each} 
parameter. We determine the associated 68\%
Bayesian confidence intervals by constructing the cumulative 
probability function for each parameter and finding the 
parameters associated with values of 0.16 and 0.84 in the
cumulative function. We then obtain an optimal set of parameters 
for each model, as well as an associated likelihood. 

\subsection{Abundances in the UFD vs the Inner Halo and the Classical dSphs}\label{mcmc_results}

High$-$resolution studies have shown that local Milky Way halo 
stars (mostly belonging to the inner halo) have approximately 
constant \xfe\, abundance ratios in the [Fe/H] range sampled by 
our data, \mbox{$-$3.5 $<$ [Fe/H] $<$ $-$1.0} \citep[e.g.,][]{McWilliam95,Cayrel04,Cohen04}. 
Figure~\ref{fig: alphafe_ufds_all} shows that the halo \xfe\, is indeed flat in all four elements 
in our [Fe/H] range. In contrast, the classical dSphs are known to have lower
\alphafe\, at $\rm{[Fe/H]}\gtrsim{-2}$, while their \alphafe\, abundance patters may 
broadly resemble the halo at $\rm{[Fe/H]}\lesssim{-3}$ \citep[e.g.,][]{Cohen10}.

In order to compare the UFD abundance pattern to 
the halo and the dSphs, we use the technique described in 
\S\,\ref{mcmc_ssec} to fit each of the three models to (a) a 
restricted sample of the five UFDs with ancient stellar 
populations and a trend of increasing \alphafe\, with 
decreasing [Fe/H] (denoted by filled squares in 
Figure~\ref{fig: alphafe_ufds}); (b) the full UFD
sample; and (c) the CVn~I sample. For each dataset, 
we obtain best$-$fitting Models A, B, and 
C, and the associated maximum likelihoods, $\mathcal{L_{A}\,,L_{B}\,,L_{C}}$. 
We then assess the goodness of fit between each of the best$-$fitting models. 
We note that these are nested models, such that Model A is a subset of Model B, 
itself a subset of Model C. We can thus use the likelihood ratio test in order to 
compare whether the more complex model is \textit{statistically} a better fit that the 
simpler one. We compare two models at a time. Given the \textit{best$-$fit} set of 
parameters for each of two models, e.g., A and B, the simpler model can be rejected
at the $(1-\alpha)\times100\%$ level using the inequality in Equation \ref{eqn: likelihood_test}. 

\begin{equation}
R_{\mathcal{B,A}}\equiv{}2\,ln \frac{\mathcal{L_{B}}}{\mathcal{L_{A}}} > F_{\chi}(\alpha;\,n_{B}-n_{A}) = F_{\chi}(\alpha;\,k=1)
\label{eqn: likelihood_test}
\end{equation} 

Here, $F_{\chi}(\alpha;\,n_{B}-n_{A})$, is the cumulative $\chi^{2}$ function 
with $n_{B}-n_{A}$ free parameters. In our case, $n_{B}-n_{A} = 1$. 

\begin{figure}[t!]
\centering
\includegraphics[width=.48\textwidth]{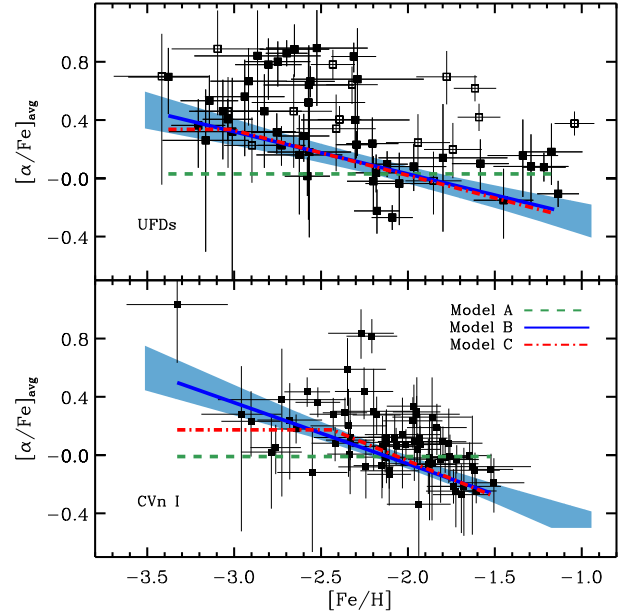}	
\caption{\textit{Top panel: } Fits to models for $\alphafeavg\,-\rm{[Fe/H]}$ trends 
for the five UFDs with ancient stellar populations and increasing \alphafe\, 
with decreasing [Fe/H] (shown with filled squares). The empty squares 
indicate stars in the three excluded UFDs (see Figure~\ref{fig: alphafe_ufds}). 
\textit{Bottom panel}: Fits to models for the CVn~I dSph. The green (dashed), blue (solid), 
and red (dot$-$dashed) lines indicate the best$-$fitting "Flat Model" (A), 
"Linear Model" (B), and "Knee Model" (C), respectively. The shaded blue 
band denotes the range of slopes within the joint 68\% confidence region 
of the two parameters in the Linear Model. In the CVn~I panel the best$-$fitting trend for 
the Knee Model is clearly distinct from the 68\% band of the Linear Model. 
Using a likelihood ratio test applied to both models, we show that the Knee Model 
is significantly preferred over the Linear Model in the case of \alphafeavg\, for CVn~I
The same test cannot distinguish between the Linear and Knee Models when 
applied to the combined UFD sample, even for \alphafeavg. The results
are unchanged if using all eight UFDs. Thus, the UFD sample is 
consistent with rising \alphafeavg\, towards lower [Fe/H], but its 
abundance pattern may differ from that of CVn~I at $\rm{[Fe/H]}\lesssim{-2.5}$ 
and below. This suggests that chemical evolution is not universal at the faintest galaxy 
luminosities, since brighter systems, e.g. CVn~I, appear to have a knee at higher 
[Fe/H] than the UFDs. The results of the likelihood ratio test are provided in 
Table \ref{table: mcmc}.}
\label{fig_models_grid}
\end{figure}

We report the likelihood ratio ($\mathcal{R}$ values) in 
Table \ref{table: mcmc}. The best$-$fit models for the 
restricted UFD sample and the CVn~I sample in the 
\alphafeavg\,$-$[Fe/H] plane are presented in 
Figure \ref{fig_models_grid}. In both the CVn~I and 
UFD panels, the blue band represents the range of 
slopes consistent within the \textit{joint} 1$\sigma$ 
uncertainty contour of $m$ and $y_{0}$. 

We first ask whether the UFD population has an abundance 
pattern consistent with the flat inner halo, using both the
restricted and the full UFD sample. The Flat model can be 
ruled out at the 90\% (99.5\%) level if $\mathcal{R_{B,A}} \geq +2.7$ (+7.9). 
We measure $\mathcal{R_{B,A}} = 6.02$ for [Ca/Fe] in the UFD 
restricted sample (+2.50 for the full sample), and 
$\mathcal{R_{B,A}} > 10$ for [Si/Fe], [Ti/Fe], and \alphafeavg\,
for both samples, thus strongly ruling out the Flat model. 
This is also evident from a visual inspection of Figure \ref{fig_models_grid} 
in the case of \alphafeavg. We have noted that only \nmg\, stars 
have [Mg/Fe] measurements, and these are not evenly distributed 
among all UFDs. Hence we do not regard this fit as significant. 
We also perform a comparison of the Flat and Linear models for 
CVn~I, and similarly conclude that the Linear model is a better 
fit than the Flat Model for all abundance ratios ($\mathcal{R_{B,A}}$ 
ranges from +5.78 to +48.66). \textit{Hence, both UFDs and 
brighter dSphs have alpha abundance patterns different than the 
Milky Way inner halo.}

Next, we test the UFD sample and CVn~I for 
the influence of Type Ia SNe enrichment by 
comparing the Linear Model against the Knee Model, 
which has one more free parameter. Again, we perform
this test for the five UFDs with a clear \alphafe\, trend, and
for the full UFD sample. In both cases, the likelihood 
ratio test indicates that the UFD data
is consistent with the Linear model (within the range of [Fe/H] 
of our data), so that adding a knee does not improve the fit. 
In contrast, we find that that the CVn~I \alphafeavg\, data is 
best$-$fit by a Knee model with a knee at \mbox{[Fe/H] $= -2.45^{+0.13}_{-0.21}$ dex}. 
The [Fe/H] value of the knee in the \alphafe, and [Si/Fe] plots 
agree within their 1$\sigma$ uncertainties. While the $\mathcal{R}_{C,B}$
value for \alphafeavg\, suggests that a knee is present at this low 
[Fe/H], our data cannot rule out a model without a knee in the case of 
the individual alpha elements, for which $\mathcal{R}_{C,B} < 2.2$.  
The hint of a knee in CVn~I at \mbox{[Fe/H] $> -2.5$} suggests that chemical 
evolution may not be a uniform process across dwarfs with different 
luminosities, since the population of fainter UFDs does not 
have a knee at \mbox{[Fe/H] $\gtrsim{-3}$}.

We caution that the crudeness of these toy models 
means that the evidence for a knee and \textit{flat} abundance trend 
at lower [Fe/H] should be better interpreted broadly as evidence 
for a change in behavior in the \alphafe$-$[Fe/H] plane indicative of 
the onset of Type Ia SNe, and not as strictly "flat" [$\alpha$/Fe] ratios 
at low metallicities. Even in the absence of Type Ia SNe, chemical 
enrichment likely depends on the mass of the progenitor Type II~SNe 
\citep{Woosley95,Nomoto06}. If the number of Type~II~SNe is small, then the first 
(most massive) SNe enrich the ISM with higher alpha abundance ratios, 
which then decrease as less massive SNe explode. This can result in
a non$-$zero negative slope in \alphafe$-$[Fe/H], if the UFDs metallicity
increases with time.

\begin{deluxetable*}{ccccccc}
\tablecolumns{7}
\tabletypesize{\footnotesize}
\tablecaption{Relative Statistical Likelihood for Halo, Linear, and Knee Models}
\tablehead{
\colhead{Element} & 
\colhead{$R_{\mathcal{B,A}}$ (UFDs$^{a}$)} &
\colhead{$R_{\mathcal{C,B}}$ (UFDs$^{a}$)} &
\colhead{$R_{\mathcal{B,A}}$ (UFDs$^{b}$)} &
\colhead{$R_{\mathcal{C,B}}$ (UFDs$^{b}$)} &
\colhead{$R_{\mathcal{B,A}}$ (CVn~I)} &
\colhead{$R_{\mathcal{C,B}}$ (CVn~I)} 
}
\startdata
$[\alpha$/Fe$]$ & +29.82 & $-$0.20 & +10.70 & $-$0.36 & +48.66 & +4.10 \\[+0.05in]
$[$Mg/Fe$]$ & $-$0.72 & $-$0.18 & $-$0.68 & $-$0.24 & +5.78 & $-$0.72 \\[+0.05in]
$[$Si/Fe$]$ & +20.68 & $-$0.36 & +36.56 & +0.18 & +28.18 & +2.24 \\[+0.05in]
$[$Ca/Fe$]$ & +6.02 & $-$0.10 & +2.50 & $-$0.14 & +9.96 & +1.40 \\[+0.05in]
$[$Ti/Fe$]$ & +24.04 & +0.08 & +22.58 & $-$0.16 & +40.06 & +0.54
\enddata
\renewcommand{\thefootnote}{\alph{footnote}}
\footnotetext[1]{Restricted UFD sample: Systems with ancient stellar populations \textit{and} qualitative trends of increasing \alphafe\, with decreasing [Fe/H]: Com~Ber, CVn~II, Leo~IV, UMa~I, and Herc}
\footnotetext[2]{All UFDs}
\label{table: mcmc}
\tablecomments{Comparison of likelihood of best$-$fit parameters 
for the models discussed in \S\,\ref{mcmc_results}. Equation 
\ref{eqn: likelihood_test} defines $R_{x,y}$ for any two models 
$x$ and $y$. The letters A, B, and C stand for "Flat Model", 
"Linear Model", and "Knee Model". The best model fits for the UFDs$^{a}$
and CVn~I are presented in Figure~\ref{fig_models_grid}.}
\end{deluxetable*}

\subsection{Anomalous Abundance Ratios}\label{ssec_anomalous}

Anomalous abundance ratios may be associated 
with enrichment from individual Type~II SNe, given 
the mass$-$dependence of Type~II SNe ejecta \citep[e..g,][]{Nomoto06}. 
Recent papers \citep[e.g.,][]{Koch08,Feltzing09} have 
reported a few stars with anomalous abundance ratios,
but the presence of such stars remains controversial
(e.g., \citealt{Gilmore13} do not confirm the measurement by
\citealt{Feltzing09}). In the context of the UFDs studied 
here, \citet{Koch08} reported a [Mg/Ca] = +0.94 star 
(their Her$-$2 object), and estimated its abundance 
pattern could be matched by the ejecta of a
\mbox{$\sim{35} \rm{M}_{\odot}$} Type~II SN. We 
search our sample for anomalous [Mg/Ca] and [Si/Ca] 
abundance ratios. We conservatively define 
an abundance ratio between two alpha elements ([X/Y])
as anomalous if (a) the abundance ratio is more 
than 1$-\sigma$ greater than $+0.5$ or 1$-\sigma$ less than $-0.5$, 
\textit{and} (b) the abundance ratio is discrepant by more than 1$-\sigma$ from the 
mean computed for the entire UFD sample, \mbox{$\langle\rm{[X/Y]}\rangle$}.

We tentatively identify anomalous abundance 
ratios in two stars, one of which is the object 
reported in \citet{Koch08}. Our measurement
for that star is [Mg/Ca] $=+0.72\pm{0.21}$ at 
\mbox{$\rm{[Fe/H]} = -2.09\pm{0.12}$}. The 
mean [Mg/Ca] for the subsample with both 
[Mg/Fe] and [Ca/Fe] measurements is 
\mbox{$\langle\rm{[Mg/Ca]}\rangle= +0.32\pm{.07}$} 
(error on the mean). We caution that only 10 stars have 
a [Mg/Fe] measurement, due to the weakness of 
Mg spectral features, making this subsample small. 
We also identify in UMa~II a single star 
\mbox{([Fe/H]$=-2.90\pm{0.16}$)} with 
an anomalous [Si/Ca] abundance ratio 
\mbox{($\rm{[Si/Ca]}=1.17\pm{0.37}$)}, where
the mean [Si/Ca] ratio for our sample is 
\mbox{$\langle\rm{[Si/Ca]}\rangle= +0.20\pm{.05}$}.
This star is also studied by \citet{Frebel10b} as
UMa~II$-$S2, but they only report an upper limit 
on [Si/Fe]. Using their [Si/Fe] upper limit and their 
[Ca/Fe] measurement, we obtain \mbox{$\rm{[Si/Ca]} < +1.08$},
consistent with our measurement. In total, we measure 
a moderately anomalous [Si/Ca] ratio in 1 out of 26 stars, 
and an anomalous [Mg/Ca] ratio in 1 out of 10 stars. Our 
results suggest that the fraction of stars with
anomalous ratios of two alpha elements is small. 

\section{Discussion}\label{disc_sec}

\subsection{Chemical Evolution in Individual UFDs}\label{disc_indiv}

The distribution of alpha abundances allows to us to 
build a general picture of chemical evolution in the UFDs. 
The first explosions from Population III (Pop III) and/or 
massive Pop II stars provide the initial chemical enrichment 
of the UFD's ISM, depositing large quantities of alpha 
elements into the gas from which later stars form. The low 
[Fe/H], high \alphafe\, stars in our sample are formed from 
metal$-$enriched gas from these early SNe. This is a 
general feature in all of our UFDs. In contrast, our sample 
includes UFDs with and without low \alphafe\, abundance 
ratios at high [Fe/H]. We note that the two dwarfs without
low \alphafeavg\, both have M$_{V} > -4.2$. Hence, this 
distinction may still hint at a threshold for significant chemical 
evolution at M$_{V} \sim -4$. However, Com~Ber also has
M$_{V} > -4.2$ but does contain low \alphafeavg\, stars. 
In addition to the general trend with [Fe/H], we note
that there is a hint for an increase in scatter (beyond 
the observational uncertainties) in \alphafeavg\, 
at lower [Fe/H]. This may hint towards inhomogeneous 
chemical enrichment \citep[e.g.,][]{Argast00,Oey00}, 
where the products of individual SNe contaminate
different regions of the ISM without complete mixing.
We discuss the two different abundance patterns in turn, 
but strongly caution that  this distinction may only be the 
result of small samples for individual UFDs. 

\subsubsection{M$_{V} < -4.2$ UFDs and Com~Ber}\label{sssec_decreasing}

Every bright UFD in our sample with M$_{V} <-4.2$ 
(CVn~II, Leo~IV, UMa~I, Herc, and Leo~T) 
contain stars with solar or subsolar \alphafe\, 
abundance ratios, the majority of which cluster 
at high [Fe/H]. The presence of low \alphafe\, 
ratios is a strong indicator that star formation 
proceeded at least as long as needed for 
Type~Ia SNe to explode. 
The minimum time delay between the onset 
of star formation and the first Type~Ia SNe, 
$t_{\rm{min,Ia}}$, is not yet constrained 
precisely, but may be as short as $\sim{\tsnemin}$ 
Myr (see review by \citealt{Maoz12a}). In our discussion,
we adopt \mbox{$t_{\rm{min,Ia}} = 100$ Myr}. Thus, the low 
\alphafe\, stars suggest that star formation in these 
UFDs lasted longer than $t_{\rm{min,Ia}}$, and 
that the first generation of Type~II SNe does not 
succeed in quenching star formation in these systems. 
Additionally, the decrease in \alphafe\, 
ratios at $\mbox{[Fe/H]}\sim{-2.5}$ suggest a
very small level of self$-$enrichment prior to
the onset of Type~Ia SNe. This is consistent 
with a system with a very low star formation rate. 

\subsubsection{Seg~1 and UMa~II (M$_{V} > -4.2$)}\label{sssec_nondecreasing}

The Seg~1 and UMa~II UFDs are the only systems 
which do not show any low \alphafe\, stars. This suggests 
that star formation lasted less than $\sim\tsnemin\,\rm{Myr}$. 
In contrast to all other UFDs, the high \alphafe\, stars 
extend to much higher [Fe/H]. Such high [Fe/H] stars 
are difficult to explain using the trends found for the 
other UFDs, i.e. systems with low star formation efficiencies. 
\citet{Frebel12} have recently described a scenario for 
such \textit{one$-$shot chemical enrichment}. In this 
picture, after the first Pop~III stars, there is a single 
epoch of star formation, after which all gas is blown 
out of the system by SNe feedback or reionization.  
In this picture, the large spread in [Fe/H] arises from 
highly inhomogeneous gas mixing, as also pointed 
out by \citet{Argast00} and \citet{Oey00}. Large [Fe/H] 
spreads can perhaps instead be the result of accretion of 
multiple progenitor systems, each with a different 
metallicity \citep{Ricotti05}. An alternative explanation 
for the early loss of gas before $t_{\rm{min,Ia}}$ 
is gas stripping due to accretion onto the Milky Way. 
We thus consider in more detail the different possibilities 
for the quenching of star formation, taking into account 
our \alphafe\, results for each UFD.

\subsection{Quenching of Star Formation}\label{disc_quench}

We saw above that UMa~II and Seg~1 show hints of
truncated star formation. If this is due to 
gas stripping due to accretion into the Milky Way, 
then the time of accretion into the halo should 
be comparable to the time of star formation 
quenching. Recently, \citet{Rocha12} have studied the 
relation between time of infall and present$-$day 
galactocentric position and line$-$of sight velocities 
of subhalos in the Via Lactea~2 simulation, calculating 
the probability distribution of the infall time for UFDs 
and classical dSphs (their Figure 4). Interestingly, 
only Seg~1 and UMa~II, the two UFDs with flat
\mbox{$\alphafe-$[Fe/H]} patterns, show a significant 
probability of infall onto the Milky Way halo more 
than 12 ago (Com~Ber may have an infall time as 
old as $\sim{11}$ Gyr ago). The early infall suggests that
gas stripping and/or heating due to accretion played 
a role in terminating star formation before low 
$\alphafe$ stars could form. However, the presence 
of high [Fe/H] stars poses a problem to this interpretation, 
since the presence of high \alphafe, high [Fe/H] stars is 
attributed to systems with high star formation efficiencies. 
Thus, it is also possible that internal effects, e.g., winds 
from SNe, managed to get rid of or heat all of the remaining 
cold gas reservoir.  

In contrast to Seg~1 and UMa~II, the other UFDs have 
inferred infall times younger than 10 Gyr \citep{Rocha12}. 
\citet{Brown12} have reported upper limits on 
the duration of star formation of less than $\sim{2}$ Gyr 
for three UFDs (UMa~I, Herc, Leo~IV), implying 
star formation was terminated in the first few Gyr 
after the Big Bang. \citet{Okamoto12} also report 
a lack of a significant age spread in CVn~II. 
Therefore, gas stripping due to accretion likely 
occurred only after quenching of star formation, 
thus limiting this process's role in the chemical 
evolution of the majority of UFDs. 

The presence of low \alphafe\, abundance ratios 
suggests that star formation lasted for at least 
\mbox{$\sim$\tsnemin\, Myr} in most UFDs. Assuming 
that the first Pop~III stars form $\sim{180}$ Myr 
($z\sim{20}$) after the Big Bang \citep[e.g.,][]{Abel00}, 
the end of star formation likely occurred \textit{after} 
$z\sim{14.5}$. This approximate upper limit on 
the redshift at which quenching occurred changes 
depending on the actual minimum time delay for 
Type~Ia SNe. If the minimum time delay for Type~Ia SNe is
on the order of 500 Myr instead (five times the value adopted
above), then quenching happened no earlier than $z\sim{7.7}$. 
The process of reionization likely extends for an extended 
range in redshift: \mbox{$14\lesssim{z}\lesssim{6}$} \citep{Fan06}.
Our data are thus consistent with either internal 
evolutionary processes, i.e. Type~II SNe blowing 
out the gas or providing enough thermal feedback 
to suppress the formation of additional stars, or 
reionization$-$driven quenching \citep{Bullock00,Ricotti05}.
In both scenarios, the effect of the first SNe 
explosions (from Pop III and/or massive Pop II stars) 
must still allow for star formation to proceed long 
enough for Type Ia SNe to explode and enrich the ISM.

\subsection{The UFDs and the Halo}\label{disc_all}

In \S\,\ref{mcmc_results},  we characterized the distribution 
of \alphafe\, abundances as a function of [Fe/H]. We showed 
that the probability that this distribution is drawn from a flat, 
inner$-$halo-like abundance pattern is less than a few percent. 
This fully agrees with a picture where the bulk of the inner 
halo forms from the accretion of a few massive satellites 
undergoing efficient star formation \citep{Robertson05}, rather
than the UFDs.

The fractional contribution of stars to the halo from UFDs may rise 
towards lower metallicities. The Milky Way halo is known 
to host extremely metal$-$poor stars (EMPs) with 
\mbox{$\rm{[Fe/H]}\leq{-3}$}. In contrast to the classical dSphs, the 
UFDs host a significant fraction of EMPs. Our sample, which is 
not metallicity$-$biased, has \nfehltthree\, EMP stars out of \nstars. 
At these low [Fe/H], the similarity in alpha enhancement between 
the halo and the combined UFD sample alone does suggest a 
larger contribution of the UFDs or UFD$-$like systems to the 
low metallicity tail of the inner halo. Since star formation likely 
ceased in the UFDs significantly prior to being accreted into 
the Milky Way potential, it suggests that the present$-$day UFD
abundance patterns may be similar to those of UFDs accreted
in the past, so that a large population of UFDs with similar 
abundance patterns may have contributed to the build$-$up of the
EMP inner halo. This picture will need to be refined and tested by
comparing other species with different nucleosynthetic origins, such as 
neutron$-$capture elements. For example, high$-$resolution studies of 
small numbers of stars in the UFDs suggest that the mean [Ba/Fe] 
ratio in Com~Ber \citep{Frebel10b} is lower than in UMa~II
or Bo\"{o}tes~I \citep{Gilmore13}. In contrast to Com~Ber, 
these two UFDs have [Ba/Fe] ratios more similar to those of 
the Milky Way inner halo at \mbox{$\rm{[Fe/H]}\sim{-2.5}$.}

As a final note, it is worth emphasizing that a comparison 
of UFD abundances to the outer halo may provide circumstantial 
evidence for a stronger link between these two populations. The 
few studies to date rely on kinematically selected, nearby halo stars 
consistent with outer halo membership based on their kinematics 
and/or calculated orbital motions \citep[e.g.,][]{Fulbright02,Roederer09,Ishigaki10}. 
\citet{Nissen10} in particular show a clearly distinguishable 
population of low \alphafe\, stars in the range $-1.6<\rm{[Fe/H]}<-0.4$.
These are likely outer halo stars with high eccentricities, which thus
take them as close as the Solar radius. However, orbits calculated 
to date for a few dSphs \citep[e.g.,][]{Piatek05} show that their orbits 
are not likely to pass as close as the Solar Galactocentric radius.
All UFDs are at present at least as far as $R_{\rm{GC}}\sim{28}$~kpc. 
Thus, a full understanding of the UFD/dSph$-$outer halo connection awaits 
a detailed mapping of chemical abundances in the in$-$situ outer halo.

\section{Conclusions}\label{conc_sec}

We analyze Keck/DEIMOS spectroscopy of RGB stars in eight UFDs 
using a spectral matching technique to measure and characterize the 
distribution of \alphafe\, abundance ratios. In this
paper, we report [Mg/Fe], [Si/Fe], [Ca/Fe], and [Ti/Fe], as well as 
a combined alpha to Fe abundance ratio, \alphafeavg\,, for \nstars\, stars
in these systems. We summarize our main conclusions as follows:

\begin{itemize}
\renewcommand{\labelitemi}{$\bullet$}
\item Out of seven UFDs with ancient stellar populations, five
(Coma Berenices, Canes Venatici II, Ursa Major I, Leo~IV 
and Hercules) show an increase of \alphafe\, towards lower 
[Fe/H], and low \alphafe\, ratios for their highest [Fe/H] stars, 
implying that Type~Ia SNe had enough time to pollute the ISM. 
This suggests that star formation was an extended 
process, lasting \textit{at least} $\sim{\tsnemin}$ Myr, corresponding 
to the minimum  time delay between the onset of star formation and the 
first Type~Ia SNe. Leo~T, which has a much more extended star formation
history, shows the same abundance pattern.
\item The remaining two UFDs with old populations, 
Segue~1 and Ursa Major II, show enhanced \alphafe\, ratios 
at \textit{all} metallicities, ranging from 
$-3.5 \lesssim{\rm{[Fe/H]}}\lesssim{-1.0}$. On average, 
the mean level of alpha$-$enhancement in Segue~I is higher 
than in UMa~II by $\sim{0.2}$ dex. The absence of low 
\alphafe\, stars suggests that the star formation period 
was very short, less than $\sim\tsnemin$ Myr.
\item The combined population of UFDs shows a clear 
increasing trend in \alphafe\, with decreasing [Fe/H], with 
no evidence for a plateau within our entire metallicity range. 
Although this \textit{rise} in \alphafe\, disagrees with the flat 
inner halo abundance pattern, a significant number of 
$\rm{[Fe/H]} < -2.5$ stars in the UFDs have abundance ratios consistent 
with the halo values within the uncertainties. Therefore, the UFD 
contribution to the extremely metal$-$poor star (EMP) halo 
abundance pattern may be more significant at the lowest metallicities. 
\item The abundance pattern in the UFDs shows some 
difference with respect to the brighter CVn~I dSph.
We show that, in contrast to the UFD population, there 
is a hint for a plateau at \mbox{[Fe/H]$\sim{-2.3}$} in CVn~I.
This difference suggests that the star formation efficiency in the 
UFDs was lower than in the more luminous dSphs.
\end{itemize}

We have shown based on \alphafe\, abundance ratios that most of the 
UFDs were able to retain gas and form stars long enough for the first 
Type Ia SNe to explode, and that this evolution proceeded inefficiently. 
The use of medium$-$resolution spectroscopy has been instrumental 
in providing us with large enough samples to begin to address the 
evolution of these systems.  Future studies will aim to study the details 
of this evolution by comparing the shape of the metallicity distribution 
function of each UFD to chemical evolution models. For instance, evidence 
for a rapid shutdown in star formation due to reionization may appear as 
an abrupt cutoff in the metallicity distribution function at the higher [Fe/H] 
end. Due to the sparseness of the RGBs of the UFDs, it will become 
necessary to extend the power of multiplex spectroscopic observations 
to the main sequence of the UFDs in order to obtain statistically significant 
samples to achieve this goal.

\section*{acknowledgements}

The authors would like to thank Ana Bonaca 
and the anonymous referee for helpful comments on 
the manuscript.  LCV is supported by a NSF Graduate Research Fellowship. 
MG acknowledges support from NSF grant AST$-$0908752 and 
the Alfred P. Sloan Foundation. ENK acknowledges support from 
the Southern California Center for Galaxy Evolution, a multicampus 
research program funded by the University of California Office of 
Research.

We wish to recognize and acknowledge the very 
significant cultural role and reverence that the summit of
Mauna Kea has always had within the indigenous 
Hawaiian community. We are most fortunate to have 
the opportunity to conduct observations from this mountain.

\bibliographystyle{apj}

\end{document}